\documentclass[pageno]{jpaper}
\pdfoutput=1
\usepackage[ruled]{algorithm2e}
\usepackage{multirow}
\usepackage{listings}
\usepackage{subfig}
\usepackage[normalem]{ulem}
\usepackage{amssymb,amsmath,amsfonts}
\usepackage{multirow}
\usepackage{comment}
\usepackage{graphicx}
\usepackage{subfig}
\usepackage{url}

\usepackage{pifont}

\usepackage{tipa}
\usepackage{tikz}

\usepackage{booktabs}

\newcommand{\arch}{our framework}

\newcommand{\tmp}{\ding{172}}
\newcommand{\2}{\ding{173}}
\newcommand{\3}{\ding{174}}
\newcommand{\4}{\ding{175}}
\newcommand{\5}{\ding{176}}
\newcommand{\6}{\ding{177}}

\newcommand{\rdO}{\ding{172} \& \ding{173}}
\newcommand{\wrO}{\ding{175} \& \ding{176}}

\newcommand{\shrink}{Eliminate vertical white-space}
\newcommand{\vshrink}[1]{
  \ifdefined\shrink 
	\vspace{-#1cm}
  \else
	\vspace{0cm}
  \fi
}

\usepackage{authblk}

\date{}
\begin{document}
\title{Trading Computation for Communication: A Taxonomy}

\author[1]{Ismail Akturk\thanks{akturki@missouri.edu}}
\author[2]{Ulya R. Karpuzcu\thanks{ukarpuzc@umn.edu}}
\affil[1]{Department of Electrical Engineering and Computer Science, University of Missouri, Columbia}
\affil[2]{Department of Electrical and Computer Engineering, University of Minnesota, Twin Cities}




\maketitle

\begin{abstract}
A critical challenge for modern system design is meeting the overwhelming
performance, storage, and communication bandwidth demand of
emerging applications within a tightly bound power budget. As both the time and
power, hence the energy, spent in data communication by far exceeds the energy spent
in actual data generation (i.e., computation),  (re)computing data can easily become
cheaper than storing and retrieving (pre)computed data.  Therefore, trading
computation for communication can improve energy efficiency by minimizing 
the energy overhead incurred by data storage, retrieval, and
communication. This paper hence provides a taxonomy for the computation vs.
communication trade-off along with quantitative characterization.
\end{abstract}

\section{Introduction}
\label{sec:intro}
\begin{figure}[h]
  \begin{center}
	\includegraphics[width=0.9\columnwidth]{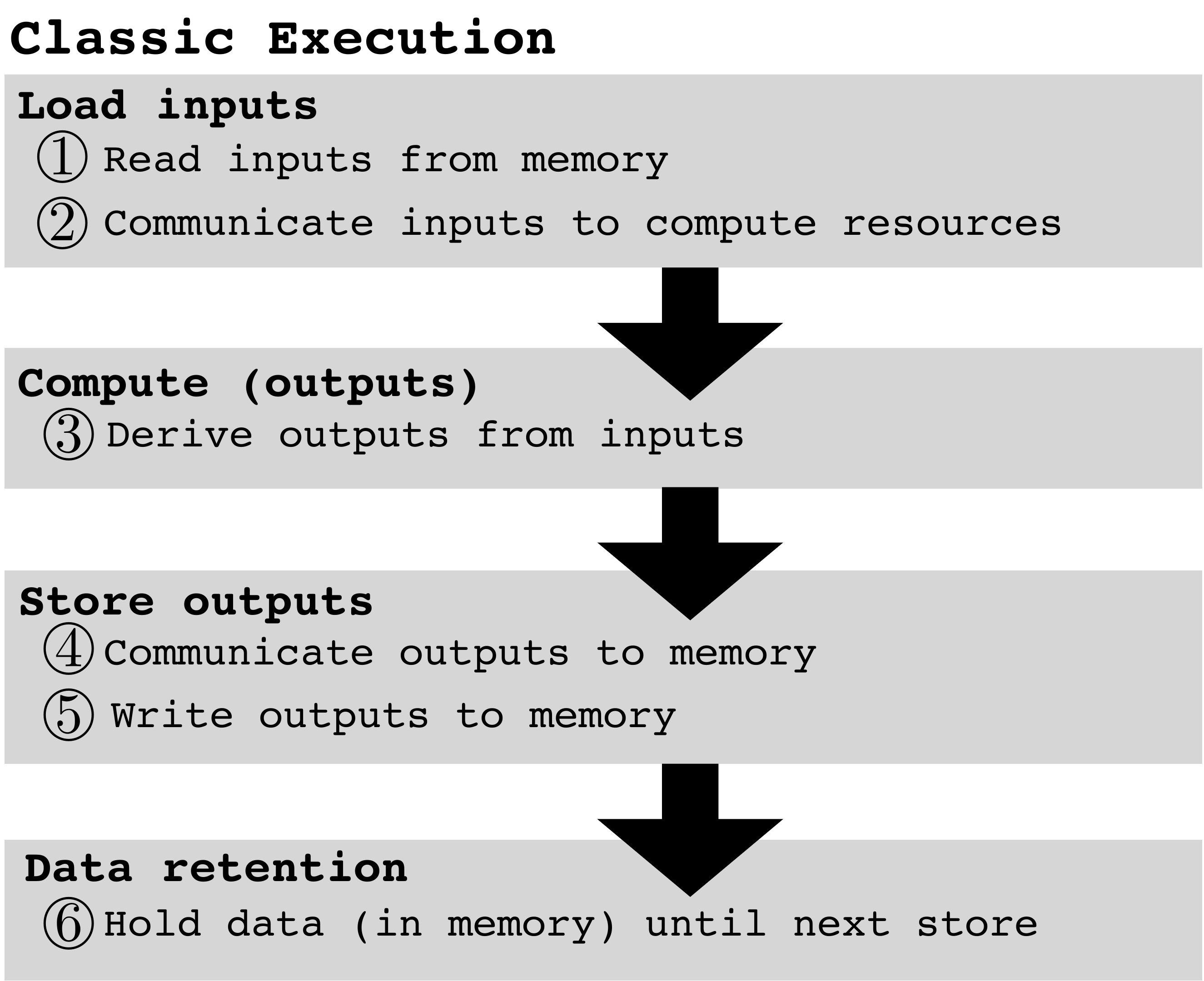}
	\caption{Classic execution at each step of computation.}
	\label{fig:st}
  \end{center}
\end{figure}

Addressing energy problem of modern computing~\cite{Horow} is not possible without
understanding {\em where the power goes}.  Figure~\ref{fig:st} demonstrates a
generic template for the sequence of events accompanying each step of classic
computing: Upon retrieval of the input operands from the memory hierarchy (\rdO), compute resources
(be it general-purpose cores or specialized accelerators) derive the output data
from the inputs (\3), followed by storage (\wrO) and retention (\6) of the
output data
until the next update.  Power goes to all of these events.  The building blocks
of classic processors, digital switches, consume dynamic power as they toggle,
and  -- being far from ideal due to aggressive miniaturization -- static power
due to leakage when turned off.  

Both the breakdown of total power consumption across events, and the ratio of
dynamic to static power per event evolve as a function of the operating regime
and technology.  Unfortunately, emerging technology solutions are not mature
enough to meet the growing performance, storage capacity, and communication
bandwidth demand within the tightly bound power budget (mainly due to cooling and power
delivery limitations).  At the same time, imbalances between logic and memory
technologies cause 
energy (time $\times$ power) consumption of data loads and stores 
(\tmp, \2, \4\ and \5)
to significantly exceed the energy consumption of 
actual computation
(\3)~\cite{Kogge,Horow}.  As a consequence, reproducing, i.e., {\em recomputing}
data can become more energy efficient than storing and retrieving
pre-computed data. 
This discrepancy is expected 
to become even more
prevalent with technology scaling~\cite{Keckler:2011kc}.
\begin{figure}[ht]
  \begin{center}
	\includegraphics[width=1\columnwidth]{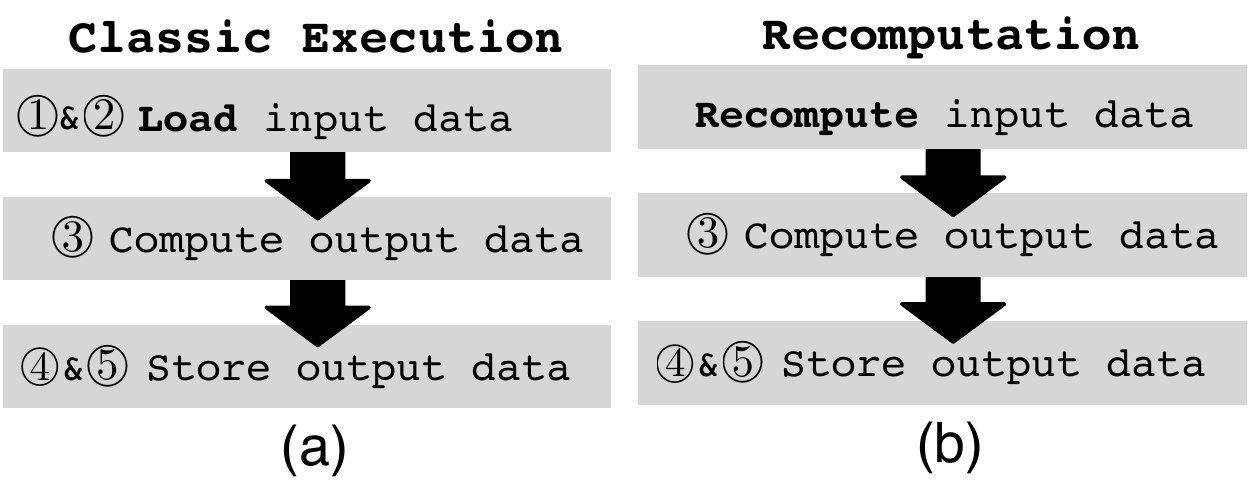}
	\caption{Classic execution vs. Recomputation}
	\label{fig:amnesiac}
  \end{center}
\end{figure}

Figure~\ref{fig:amnesiac}(a) shows the classic trajectory at each step of
execution. Black arrows point to the direction of data flow.  As depicted in
Figure~\ref{fig:amnesiac}(b), {\em recomputation} swaps load instructions for the
reproduction of the respective input operands (which would otherwise be loaded
from memory) for the subsequent computation.  \tmp\ incurs the time and power
overhead of the memory (hierarchy) access to perform the load; \2, of the subsequent
communication of inputs to compute resources.  {\em Recomputation
  transforms the overhead of \rdO\ to the overhead of the recomputation of the
respective data values, i.e., of \3}.  Therefore, recomputation can only improve
energy efficiency if the cost of data reproduction remains less than the
overhead of \rdO. 
In other words, the overhead of \rdO\ sets the budget for recomputation.

Recomputation can also reduce the pressure on memory capacity and communication
bandwidth. A recomputing processor can accommodate more compute resources (in
the form of general-purpose cores or specialized accelerators) to occupy the
area once allocated to memory (hierarchy).  At the same time, under
recomputation the workload becomes more compute-intensive to make a better use
of classic processors optimized for compute performance, as opposed to energy
efficiency.  This paper quantitatively characterizes the energy efficiency
potential of recomputation, and introduces a taxonomy for the computation vs.
communication trade-off. 
In the following, 
Section~\ref{sec:impl} introduces the taxonomy; 
Sections~\ref{sec:setup} and \ref{sec:eval} provide the evaluation; 
Section~\ref{sec:rel} covers related work, 
and Section~\ref{sec:conc} summarizes the findings.

\section{Recomputation Taxonomy}
\label{sec:impl}
\noindent The energy overhead of the load from Figure~\ref{fig:amnesiac}(a)
determines the energy budget for recomputation. Unless the energy cost of
reproducing data remains less than the energy cost of the respective
load, recomputation cannot improve energy efficiency.  Whether recomputation
can improve energy efficiency or not tightly depends on where the data reside
in the memory hierarchy -- it is the location of the data in the
memory hierarchy which determines the energy cost of the load.
On the other hand, recomputation also incurs an energy cost due to the introduction
of {\em recomputing}, {\em producer} instructions.

The taxonomy of recomputation techniques spans three dimensions.
Recomputation can reproduce the data (which otherwise would be loaded from
memory) by brute-force {\bf recalculation}~\cite{amnesiac17}, value {\bf
prediction}~\cite{Huang99,valPred}, or {\bf approximation}~\cite{approx,lva},
respectively: 
\begin{list}{\labelitemi}{\leftmargin=0.4em}
  \item  Under brute-force {\bf recalculation}, the recomputation effort goes to
  the {\em derivation of data values}, by re-executing the producer instructions
  (of the data values, which would otherwise be loaded from memory).  
  \item  Under {\bf prediction}, the recomputation effort goes to the {\em
  estimation of data values} by exploiting {\em value locality} -- the likelihood
  of the recurrence of data values~\cite{valPred} within the course of execution.  
  \item Under {\bf approximation}, the recomputation effort goes to the actual
  {\em calculation of data values} -- as it is the case for brute-force
  recalculation, however, {\em at reduced accuracy}. In this case, the compute
  resources may
  only partially execute the producer instructions (e.g., by dropping a
  subset), or perform recomputation
  on reduced-accuracy hardware. 
\end{list}
Depending on the accuracy of prediction or approximation of the data
values, {\bf
prediction} or {\bf approximation} may degrade accuracy of the end results,
which is not the case for brute-force {\bf recalculation}.  This paper focuses
on {\bf recalculation} and {\bf prediction}, and leaves {\bf
approximation} based recomputation to future work.

\subsection{Recalculation Based Recomputation}
\label{sec:recalc}
\noindent 
%
\noindent {\bf Recalculation} can be implemented in various ways. We use
compiler-based proof-of-concept implementation similar to ~\cite{amnesiac17}.
During code generation, the compiler replaces each energy-hungry load
instruction with the sequence of (arithmetic/logic) producer instructions of the
respective data values. To this end, the compiler recursively traces data
dependencies. The sequence of producer instructions forms a backwards slice, referred as {\em Recalculation Slice}, {\em RSlice}~\cite{amnesiac17}. 

\begin{figure}[htp]
  \begin{center}
	\includegraphics[width=0.75\linewidth]{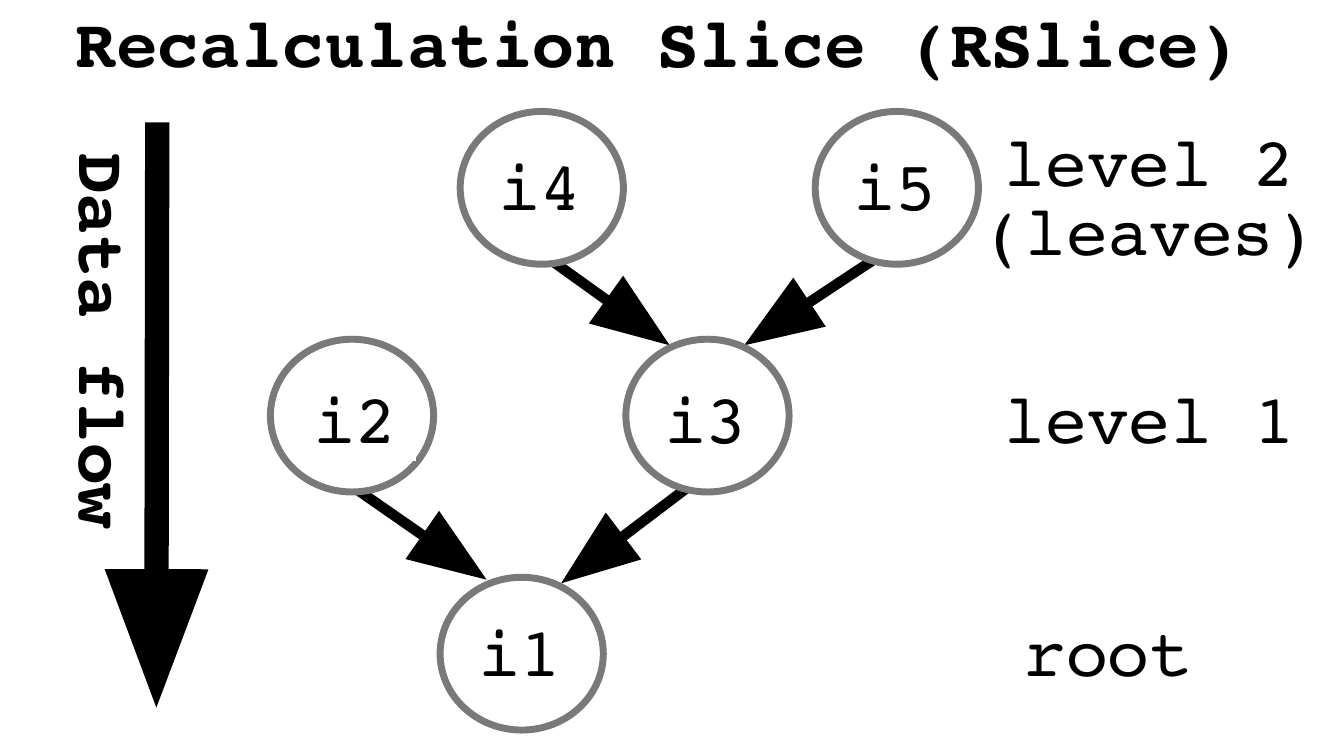}
	\caption{Example Recalculation Slice (RSlice)}
	\label{fig:rt}
  \end{center}
\end{figure}

Fig.~\ref{fig:rt} demonstrates an example RSlice. Each RSlice is an upside-down
tree, with nodes representing producer instructions to be re-executed. Data
flows from the leaves to the root. The node at the root corresponds to the
immediate producer of the data value which would otherwise be loaded from
memory. Nodes at level 1 correspond to the producers of the root.  Nodes at
level {\em l} correspond to the producers of nodes at level {\em l-1}. The
number of incoming arrows at each node reflects the number of producers (of the
node) to be re-executed.  
The leaf nodes either 
represent terminal instructions which do not have any producers, or instructions
for which re-execution of their producers is not energy efficient. In the
proof-of-concept implementation, the compiler is in charge of making sure that
all input operands of producer instructions within an RSlice are available at
the anticipated time of {\bf recalculation}. Unless the compiler guarantees
this constraint, an RSlice cannot replace its respective load in the binary. 

The compiler swaps a load with its respective {RSlice} only if {\bf
recalculation}
of the corresponding data value along the {RSlice} is more energy efficient
than performing the load. 

\subsection{Prediction Based Recomputation}
\label{sec:pred}
\noindent Under {\bf prediction}, the recomputation effort goes to the estimation of
data values, instead of brute-force {\bf recalculation}.  Accurate estimation is only
possible if data values (which otherwise would be loaded from memory) exhibit
high value locality -- i.e., a high likelihood of recurrence~\cite{valPred}
within the course of execution. 
For example, if a data value exhibits excellent (100\%) locality, just storing
the value in a dedicated buffer and retrieving it from there may turn out to be
more energy efficient than recalculating it (Section~\ref{sec:recalc}) or
loading it from memory. Even if the value locality remains less than 100\%, such
buffered history of values can be used for {\bf prediction}. It has been shown that
emerging 
applications
can oftentimes mask prediction incurred inaccuracy due to
potential errors in estimation, as implied by imperfect value
locality~\cite{valPred}. 

Value retrieval from the history buffer constitutes the main cost of {\bf
prediction}.  Under imperfect value locality, a prediction algorithm can help
estimate the respective value by using the buffered history of previously
observed values. In this case, the cost of executing the prediction algorithm
should also be considered.  The overall cost of {\bf prediction} should fit into the
recomputation budget, which in turn is set by the energy overhead of the
respective
load. {\bf Prediction} based recomputation can only be beneficial if its energy cost remains less
than the energy overhead of this load.   

\subsection{Recalculation + Prediction}
\label{sec:rPred}
\noindent {\bf Prediction} based recomputation (Section~\ref{sec:pred})
exploits locality of data values which would otherwise be loaded from memory.
With respect to {\bf recalculation} (Section~\ref{sec:recalc}), {\bf prediction} targets the value to be
produced by the root node of the RSlice.
Input values of RSlice nodes
may also
exhibit significant value locality. Let us assume that such a node {\em n} resides at level {\em
l}, and it is not a leaf. In this case, predicting {\em n}'s inputs may turn out to be
more energy efficient than re-executing {\em n}'s producers residing
at level {\em l+1} of the RSlice. Hence, combining {\bf recalculation} with {\bf
prediction}  (i.e., {\bf recalculation +
prediction}) can result in pruned RSlices to harvest even more energy efficiency.
{\bf Prediction} can also serve
identifying the inputs of leaves -- recall that, if retrieving input data of
leaves requires energy hungry memory accesses, recalculation along the RSlice
cannot be of any use.  Each intermediate node of the RSlice subject to {\bf
prediction} becomes practically a leaf, as re-execution past such nodes would no
longer be necessary.

{\bf Recalculation +
prediction} can prune RSlices,
however, even under pure  {\bf recalculation} (Section~\ref{sec:recalc}), RSlices can never grow
excessively: the energy overhead of
the respective load determines the budget for recomputation.
The cost of {\bf recalculation} increases with the number of levels, i.e., {\em
height} of the RSlice, and the number of nodes residing at each level.  The
re-execution of each node instruction incurs an energy cost.  At most, as many
nodes can be re-executed (i.e., can reside in the
RSlice) as can be fit into the recomputation budget. And {\bf recalculation} can
only improve energy efficiency if the cost of re-execution along the RSlice
remains less than the recomputation budget, which is set by the energy overhead
of the respective load.  In this manner, the energy overhead of the
load prevents excessive growth of the RSlice. 
Under {\bf recalculation +
prediction}, the cost of re-execution along the RSlice along with the cost of
selective
{\bf prediction} constitute the cumulative cost of recomputation.


\section{Evaluation Setup}
\label{sec:setup}

\noindent We experiment with benchmarks from the
SPEC2006~\cite{spec}, PARSEC~\cite{parsec}, NAS~\cite{nas}, and
Rodinia~\cite{rodinia} suites, which span emerging application domains
(Table~\ref{tbl:apps}). 
In the evaluation, we only analyze the benchmarks which harvest sizable energy
efficiency gain under
recomputation. The rest of the benchmarks did not benefit from recomputation.
The analyzed mix contains both compute- and
memory-intensive applications. Our analysis is confined to sequential,
i.e., single-threaded execution. We leave parallel recomputation to future
work.
We use a cycle accurate micro-architectural simulator, Sniper~\cite{sniper}.
The simulated microarchitecture is modeled after an in-order single-core Intel
Xeon Phi-like processor without loss of generality, which features an operating
frequency of 1.09GHz at 22nm, an L1 instruction cache of 32KB (4-way, LRU), an
L1 data cache of 32KB (8-way, LRU, WB), and an L2 cache of 512KB (8-way, LRU,
WB).
We profile the native binaries (conforming to classic execution, hence
excluding recomputation) of the benchmarks on Sniper: We record
(i) value locality of instructions at runtime (to be exploited by {\bf
prediction} based recomputation); 
(ii) cache statistics, i.e., hit and miss rates, at runtime (to derive the probabilistic
energy cost model of the compiler pass as explained in Section~\ref{sec:recalc}).

\begin{table}[ht]
\centering
\captionof{table}{Benchmarks deployed\label{tbl:apps}}
\resizebox{\columnwidth}{!}{
\begin{tabular}{||l|l|l|c||}
\hline \hline
{\bf Suite} & {\bf Benchmark} & {\bf Input}     & {\bf Application Domain} \\
\hline \hline
SPEC & 429.mcf (mcf)  & test   & Combinatorial Optimization\\
SPEC & 482.sphinx3 (sx)& test & Speech Recognition\\
NAS & is & A & Integer Sorting\\
PARSEC & canneal (ca) & simsmall & Routing Cost Minimization\\
PARSEC & facesim (fs) & simsmall & Motion Simulation\\
PARSEC & ferret (fe) & simsmall & Content Similarity Search\\
PARSEC & raytrace (rt) & simsmall & Real-time Raytracing\\
Rodinia & backpropagation (bp) & 65536 & Pattern Recognition\\
Rodinia & breath-first search (bfs) & graph1MW\_6.txt & Graph Traversal\\
Rodinia & srad (sr) & 100 0.5 502 458 1 & Image Processing\\
\hline
\hline
\end{tabular}
}
\end{table}

\noindent {\em Probabilistic Energy Cost Model for the Compiler Pass from Section~\ref{sec:recalc}:}
The 
energy per instruction
(EPI) estimates per load, store, and non-memory instructions come from measured
Xeon Phi data from~\cite{BrooksEPI}, which for memory
instructions, provides
separate EPI estimates for each level {\em Li} in the memory hierarchy: EPI$_{Li}$. 
Using these EPI$_{Li}$ and cache statistics from Sniper, we
extract probabilistic EPI estimates for loads as follows:  We derive
Pr$_{Li}$, the probability of having the load serviced by level
{\em Li}, using hit and miss statistics of {\em Li} from Sniper.  Then, the sum
of
Pr$_{Li} \times$ EPI$_{Li}$ over all levels $i$ in the memory hierarchy gives the probabilistic energy cost per load.  
Using this energy cost per load, and the EPIs for non-memory instructions, the
compiler pass swaps a load with its respective {RSlice} only if recalculation
of the corresponding data value along the {RSlice} incurs a lower energy cost
than performing the load. 

%

\noindent {\em Simulation Infrastructure:} We implement the compiler pass from
Section~\ref{sec:recalc} in a Pin~\cite{pin} based tool, which by using the
probabilistic energy cost model detailed above and by tracking data
dependencies, swaps load instructions in the binary for the respective RSlices,
only if recomputation incurs a lower energy consumption.  This tool adjusts the
binary under {\bf prediction} and {\bf recalculation+prediction} accordingly,
following Sections~\ref{sec:pred} and~\ref{sec:rPred}. We restrict
{\bf prediction} with the prediction of the values to be produced by RSlice
roots. We deploy Sniper
integrated with McPAT~\cite{mcpat} to run these annotated binaries in order to
collect performance and energy statistics under recomputation.

\section{Evaluation}
\label{sec:eval}
\noindent We next quantify the energy efficiency under recomputation and analyze the implications for execution semantics.

\subsection{Impact on Energy and Performance}
\label{sec:eval1}
\begin{figure}[t!]
  \begin{center}
        \includegraphics[width=1\columnwidth]{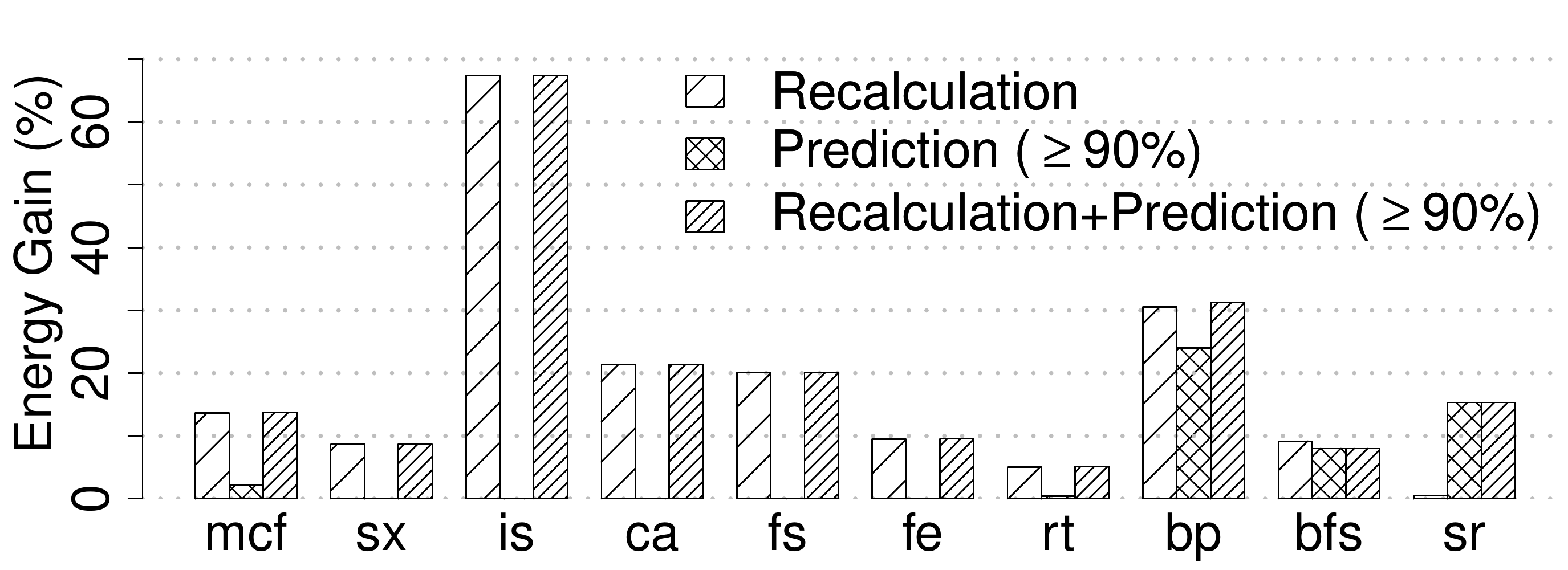}
        \caption{Energy gain under recomputation.
  \label{fig:energy}}
  \end{center}
\end{figure}
\noindent Figure~\ref{fig:energy} compares the energy consumption under {\bf
recalculation}, {\bf prediction}, and {\bf recalculation+prediction} based
recomputation.  This analysis accounts for the overhead of recomputing 
producer instructions (along RSlices) under {\bf recalculation} (Section~\ref{sec:recalc}), and history buffer
accesses under {\bf prediction} (Section~\ref{sec:pred}).
However, we assume that one history buffer access suffices for value prediction at
100\% accuracy (i.e., we omit any potential overhead due to prediction
algorithms).  For this experiment, we set the value locality threshold to
enable prediction to 90\%: prediction only applies to instructions which exhibit at least
90\% value locality.  {\bf Prediction} targets only the values to be re-produced
by {\em root} instructions of RSlices
(all instructions along which are re-executed under {\bf recalculation}).  Under {\bf
recalculation+prediction}, on the other hand, prediction can target any RSlice
instruction but the root (Section~\ref{sec:rPred}).

Figure~\ref{fig:energy} reports the energy gain with respect to native
execution, which excludes recomputation.  We observe that except bp, bfs and
sr, the energy gain under {\bf prediction} is insignificant.  This is because
only a small of number of 
instructions 
exhibit a higher value locality than 90\%.
Due to its wider applicability, {\bf recalculation} unlocks higher energy gains,
ranging from 5.06\% to 67.43\%,
except sr. The {\bf recalculation} cost for 
sr remains generally higher than the cost of the respective loads.
An interesting observation is that bfs obtains lower energy gain under {\bf
prediction} and {\bf recalculation+prediction} when compared
to {\bf recalculation} alone. The reason is that the RSlices of
bfs are very short, rendering {\bf recalculation}
always cheaper than {\bf prediction}. At the same time, our proof-of-concept
implementation gives the priority to prediction, if a value exceeds the locality
threshold set for prediction (i.e., 90\%) under {\bf
recalculation+prediction}: in other words, we omit recalculation for all values that
exhibit a higher value locality than the threshold (90\% in this case), even though
recalculation turns out to be less energy hungry than the respective load.
Therefore, the energy gain under {\bf recalculation+prediction} cannot
exceed the gain under {\bf recalculation} for bfs.
Overall, the energy gain due to {\bf recalculation+prediction} remains limited
for the majority of the benchmarks. The reason is twofold:
the benchmarks either do not have enough value locality to exploit prediction (e.g. mcf, sx, is, ca, fs, fe, and rt),
or recalculation is too costly (e.g. sr).
\begin{figure}[t!]
  \begin{center}
        \includegraphics[width=1\columnwidth]{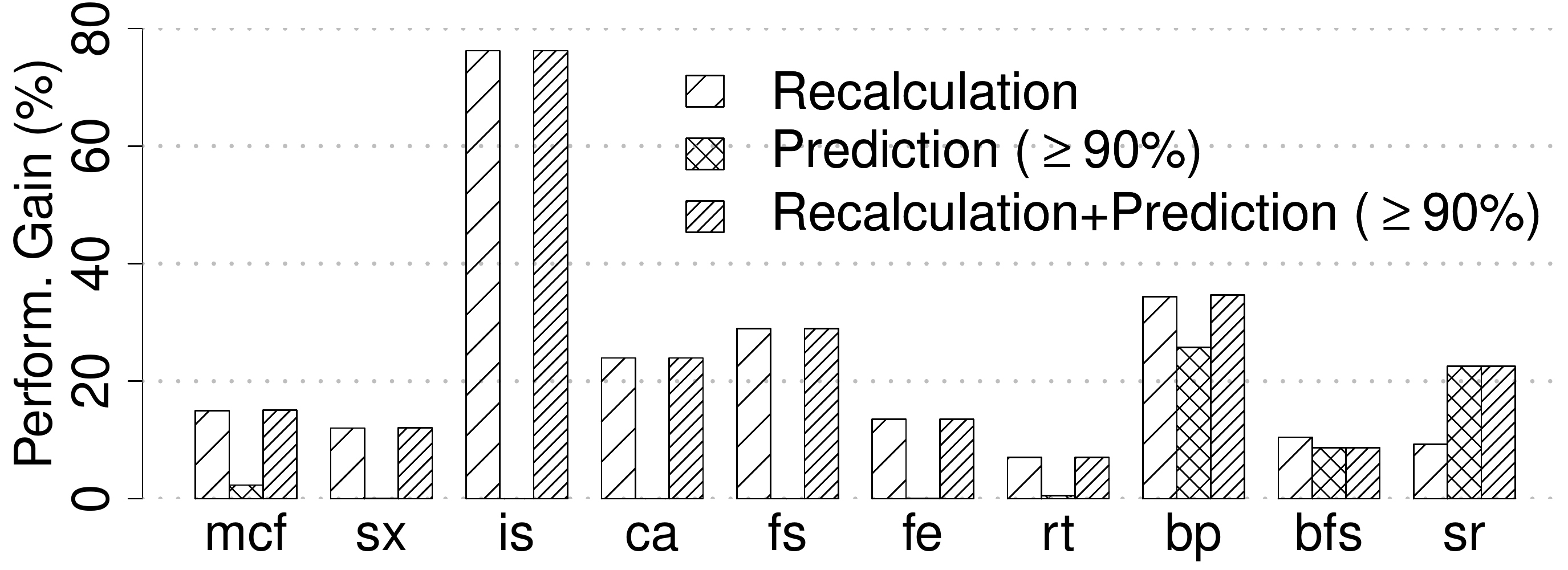}
        \caption{Performance improvement under recomputation.
  \label{fig:latency}}
  \end{center}
\end{figure}

Figure~\ref{fig:latency} reports the corresponding improvement in performance
(i.e., execution time) with respect to native execution. Generally, a similar
trend to energy gain applies, except that
%
the performance gain under {\bf recalculation} for sr becomes more pronounced
when compared to the energy gain.
\begin{figure}[t!]
  \begin{center}
        \includegraphics[width=1\columnwidth]{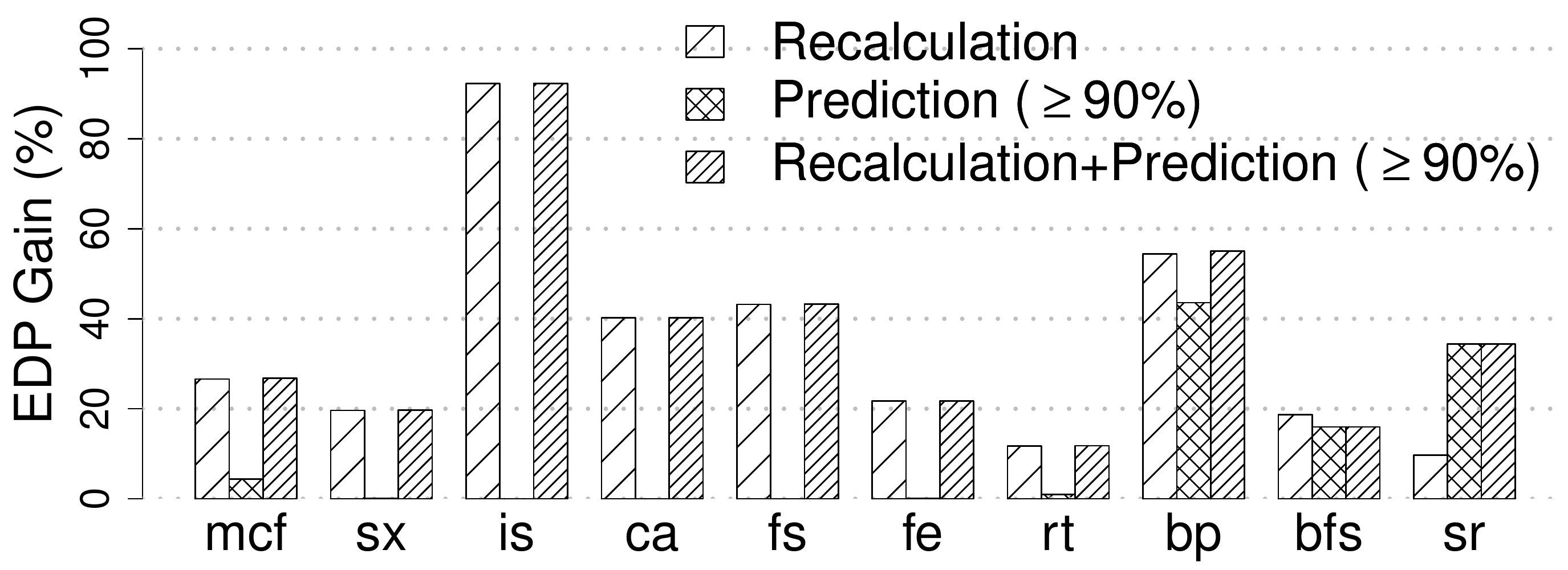}
        \caption{EDP gain under recomputation.
  \label{fig:edp}}
  \end{center}
\end{figure}

Figure~\ref{fig:edp} summarizes the resulting gain in energy efficiency in terms
of EDP (energy delay product~\cite{edp}), with respect to native execution.
Overall, {\bf recalculation+prediction} maximizes the EDP gain, and
{\bf recalculation} remains effective as well, except sr (as explained above).
{\bf Prediction} is beneficial for bp, bfs, and sr only -- recall that even this gain under {\bf
prediction} is optimistic as we neglect any algorithmic overhead. 
Finally, {\bf recalculation+prediction} results in 11.8\% to 92.2\% EDP gain across all benchmarks.
\begin{figure}[t!]
  \begin{center}
        \includegraphics[width=1\columnwidth]{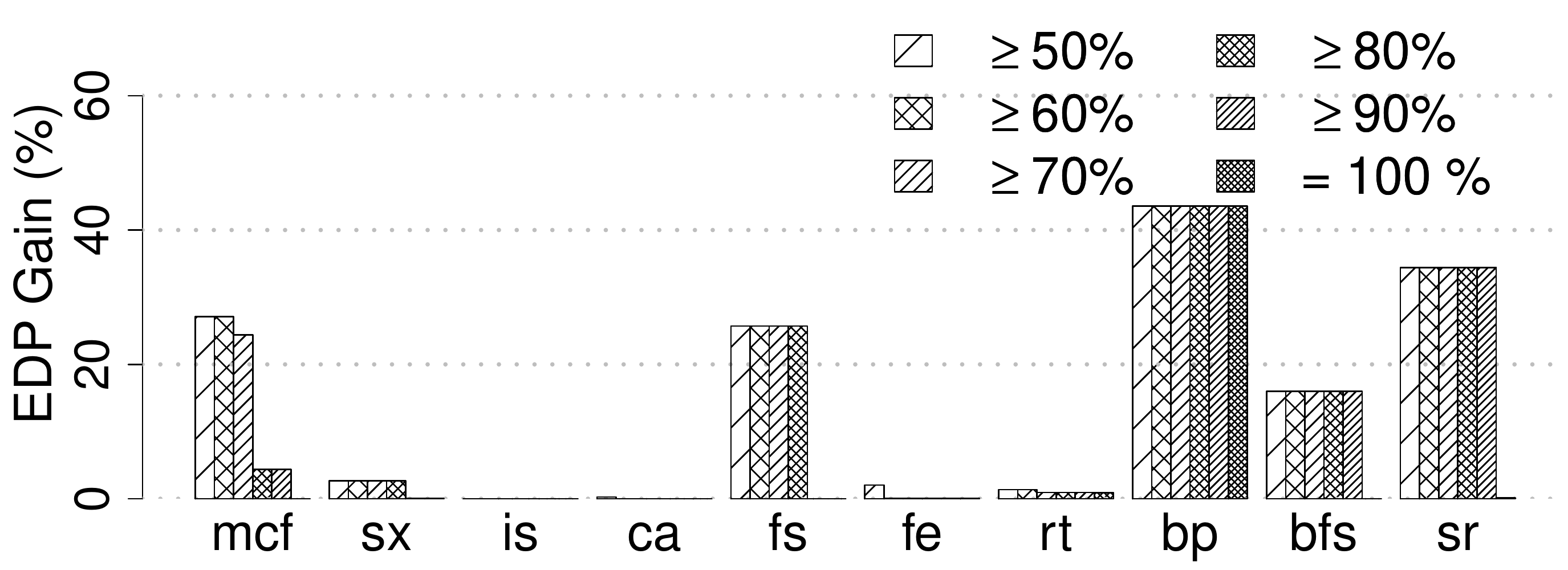}
	\caption{EDP gain under {\bf prediction} as a function of value locality threshold for prediction. \label{fig:pred_edp}}
  \end{center}
\end{figure}
\begin{figure}[t!]
  \begin{center}
	\includegraphics[width=1\columnwidth]{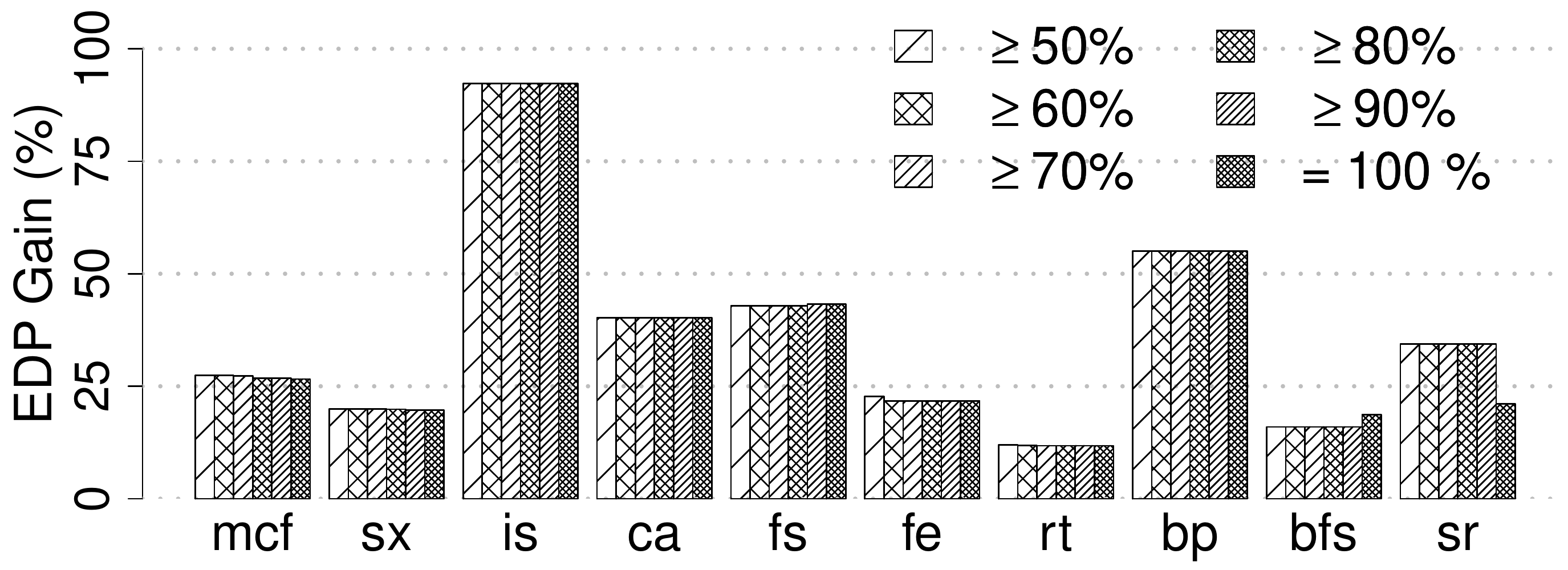}
       \caption{EDP gain under {\bf recalculation+prediction} as a function of value locality threshold for prediction. \label{fig:recandpred_edp}}
       \vshrink{0.4}
  \end{center}
\end{figure}

We next assess the sensitivity of EDP gain to the value locality threshold for
prediction.
Figure~\ref{fig:pred_edp} reports the EDP gain under {\bf prediction};
Figure~\ref{fig:recandpred_edp}, under {\bf recalculation+prediction}, as we
sweep the threshold between 50\% and 100\%.
Each bar per benchmark represents a different value locality threshold from this
range to enable prediction.  Generally, as the threshold increases, the number
of instructions exhibiting at least that much locality reduces -- therefore, a lower
number of predictions can be performed, and both the energy and performance
gains drop accordingly. 
Among the benchmarks, bp exhibits the highest value
locality, hence, it benefits most from {\bf prediction}. bfs and sr, as well,
benefit from {\bf prediction}
if the threshold remains lower than 100\% -- as very small number of loads swapped
for RSlices feature 100\% value locality for these benchmarks.
On the other hand, fs and mcf harvest sizable EDP gain under {\bf prediction} only if the threshold
remains lower than 90\% and 80\%, respectively.
The remaining benchmarks have a very small number of load instructions
that exhibit $\geq$ 50\% value locality, so only a negligible EDP gain applies under {\bf
prediction} (which already represents an upper limit for actual gains, as we neglect
any algorithmic overhead).  Therefore, {\bf recalculation+prediction} can generally provide higher EDP gains when
compared to {\bf prediction}.
As mentioned before, bfs has small RSlices, thus, the associated recalculation
cost usually remains lower than than the cost of 
prediction. Accordingly, bfs shows higher EDP gain for 100\% threshold (at which a
smaller number of values can be predicted, by definition, when compared to lower
values of the threshold) under {\bf recalculation+prediction}.
%
Overall, we observe that our findings from Figure~\ref{fig:edp} generally apply
over this wider range of threshold values. 
We can conclude that {\em recalculation has wider coverage for recomputation than
prediction}. Next, we investigate why this is the case.

\begin{figure*}[t!]
  \begin{center}
	\kern-0.8em
        \subfloat[mcf]{
                \includegraphics[width=0.25\textwidth]{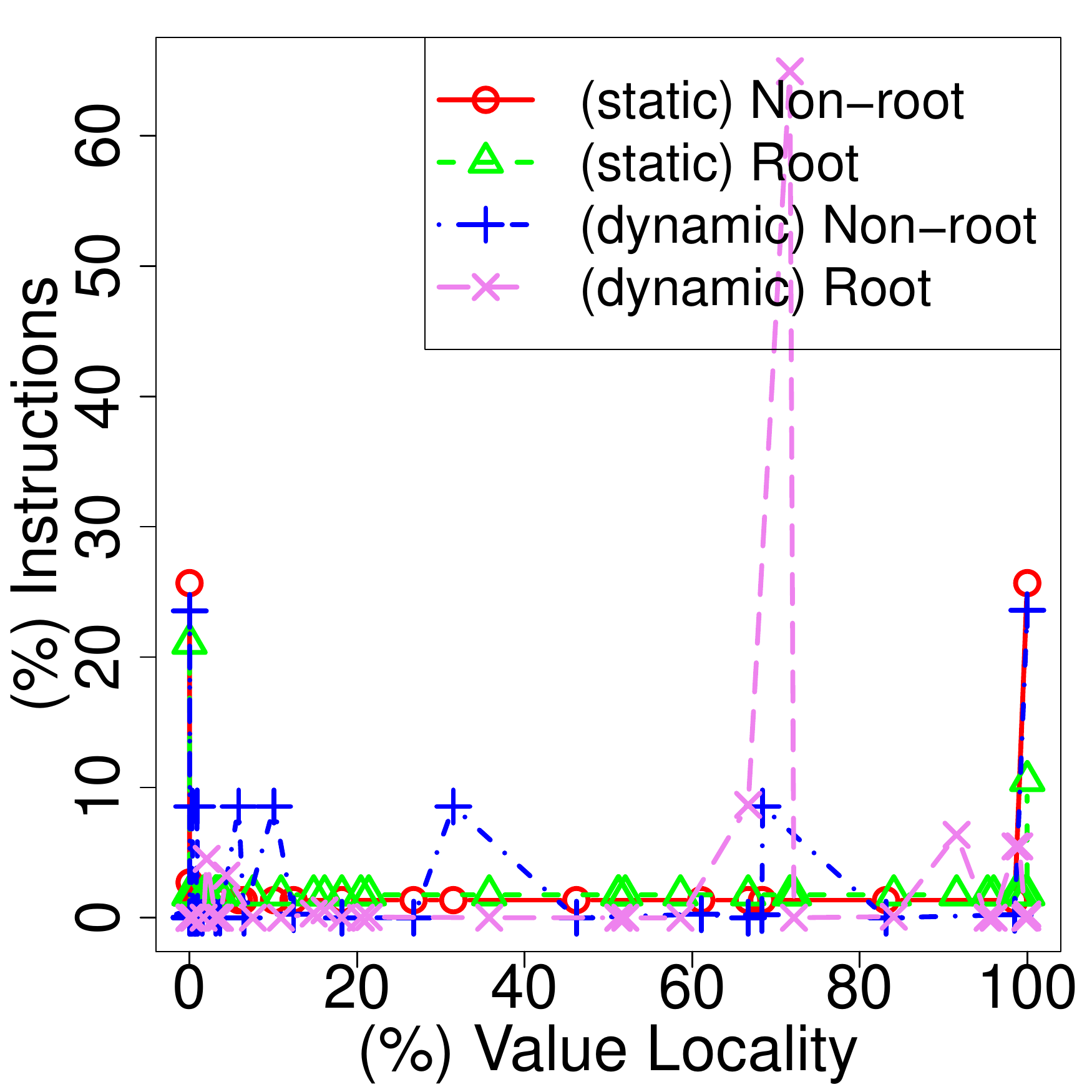}
        } \kern-0.8em
        \subfloat[sphinx3]{
                \includegraphics[width=0.25\textwidth]{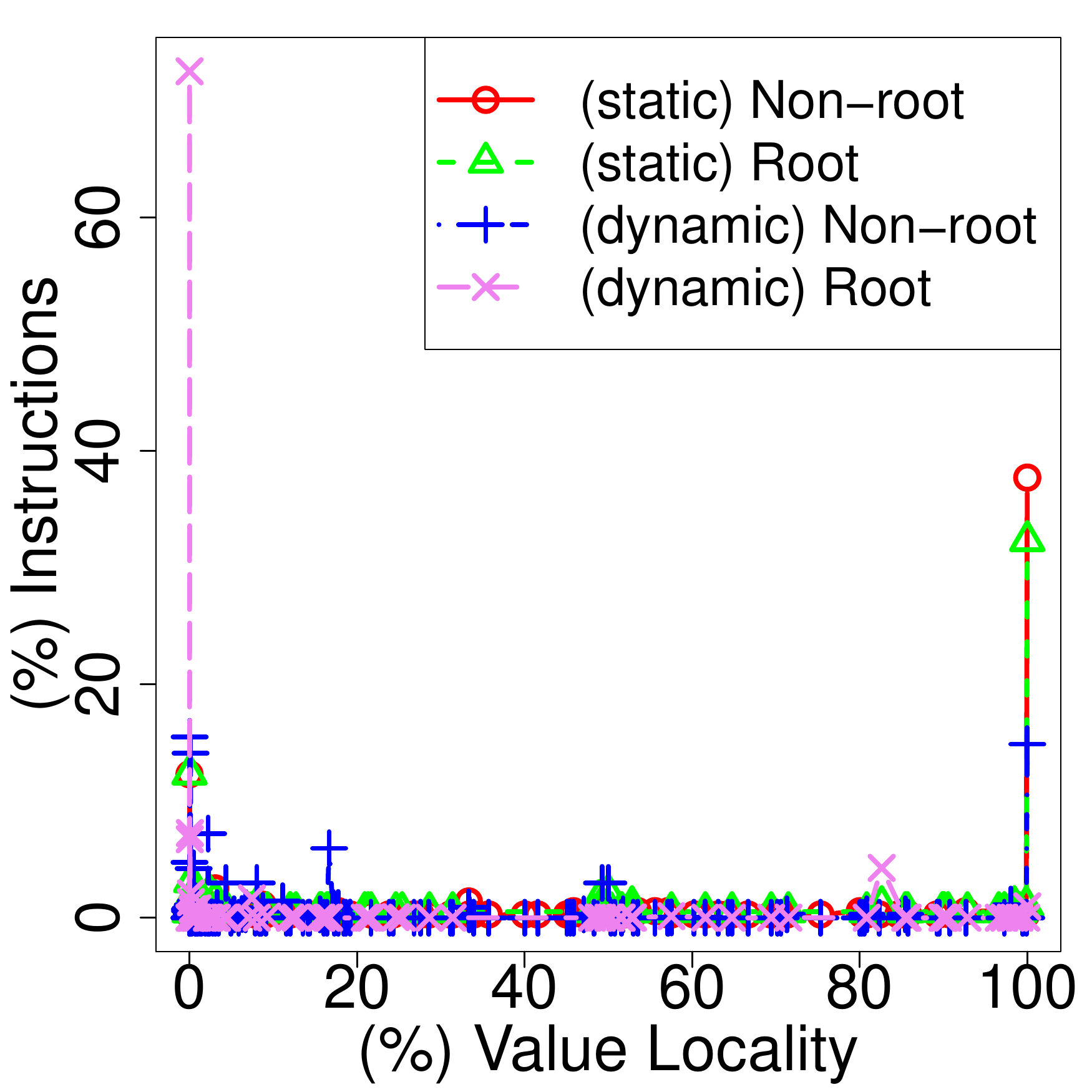}
        } \kern-0.8em
        \subfloat[is]{
		\includegraphics[width=0.25\textwidth]{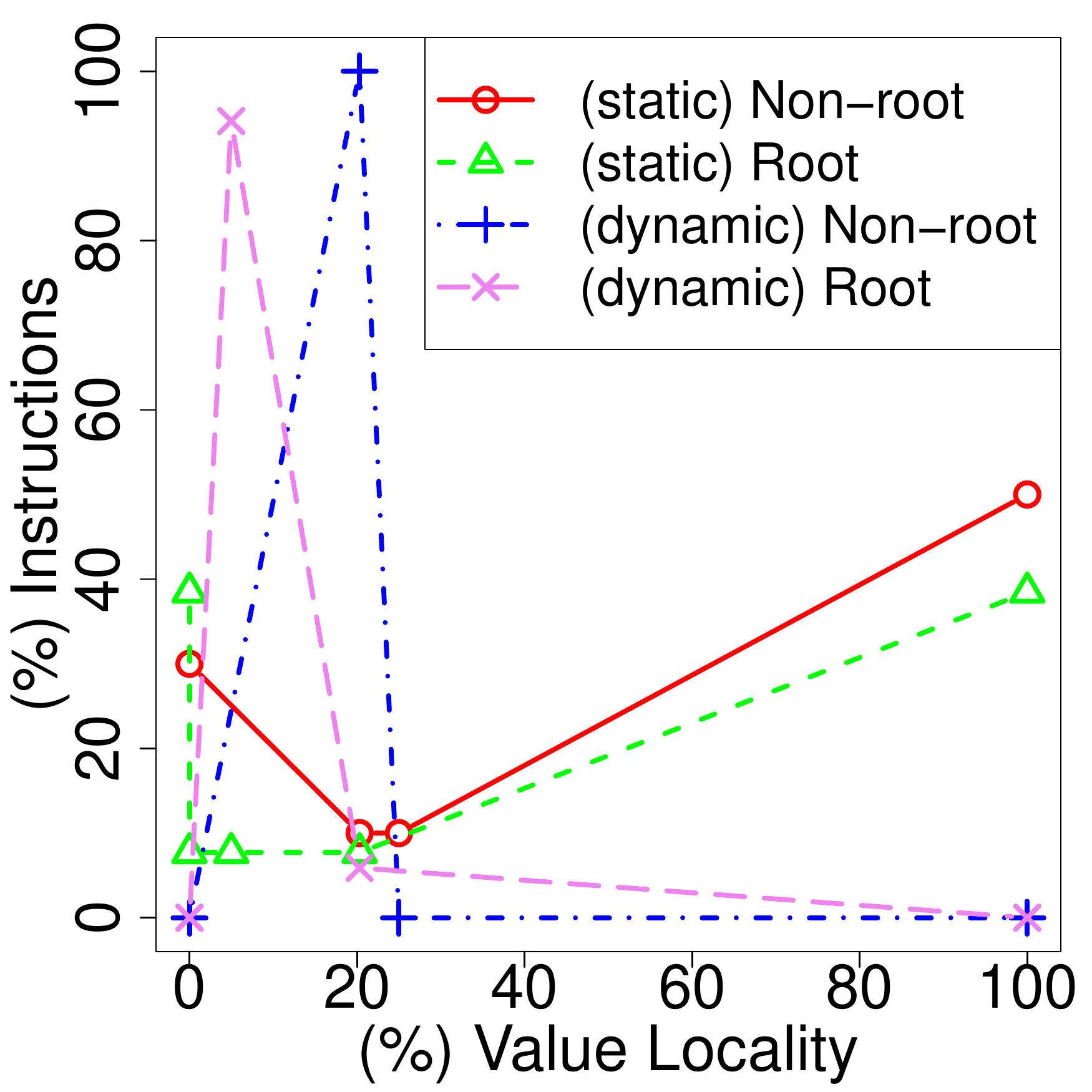}
		\label{fig:is_locality}
        } \kern-0.8em
        \subfloat[canneal]{
                \includegraphics[width=0.25\textwidth]{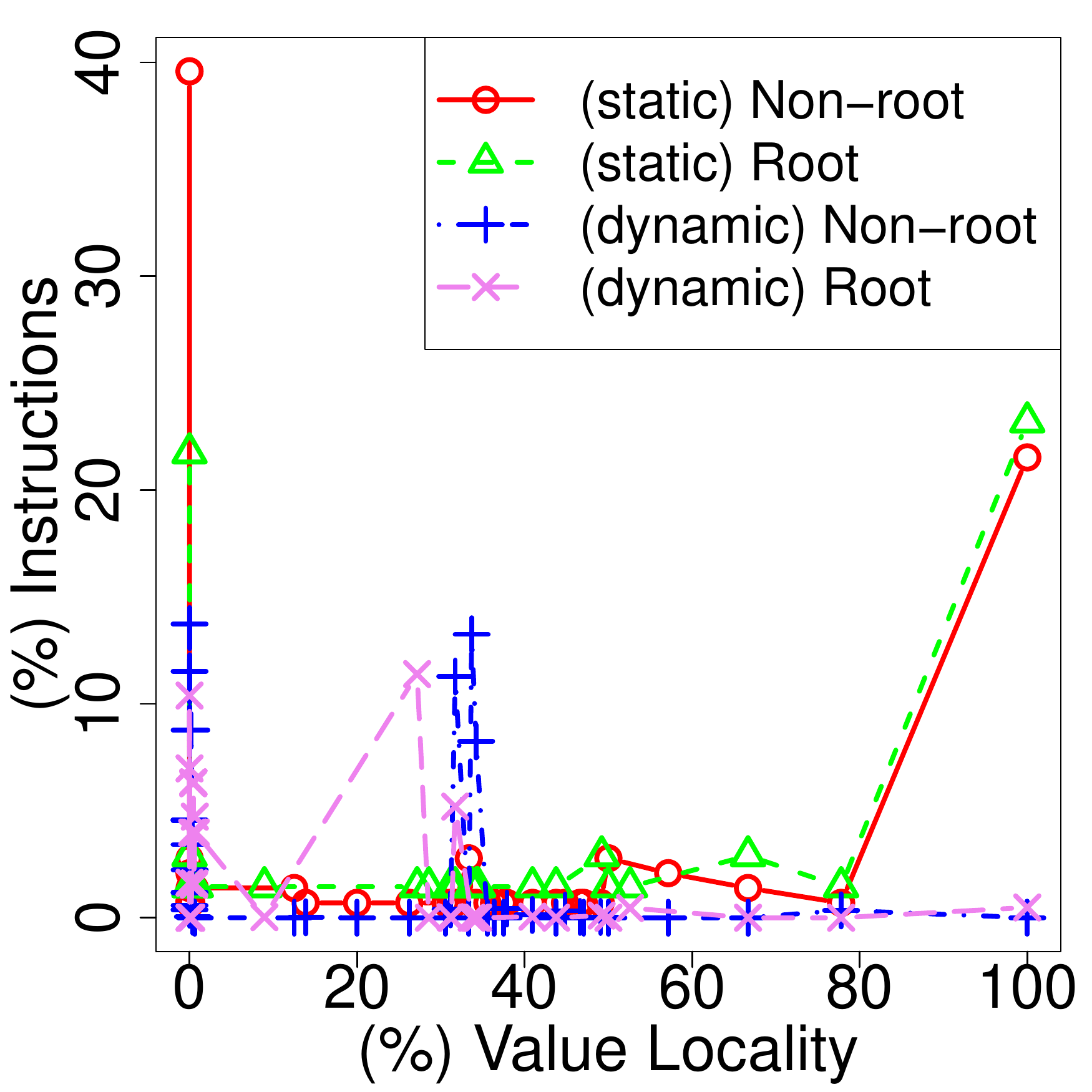}
        } \\ \kern-0.8em
        \subfloat[facesim]{
                \includegraphics[width=0.25\textwidth]{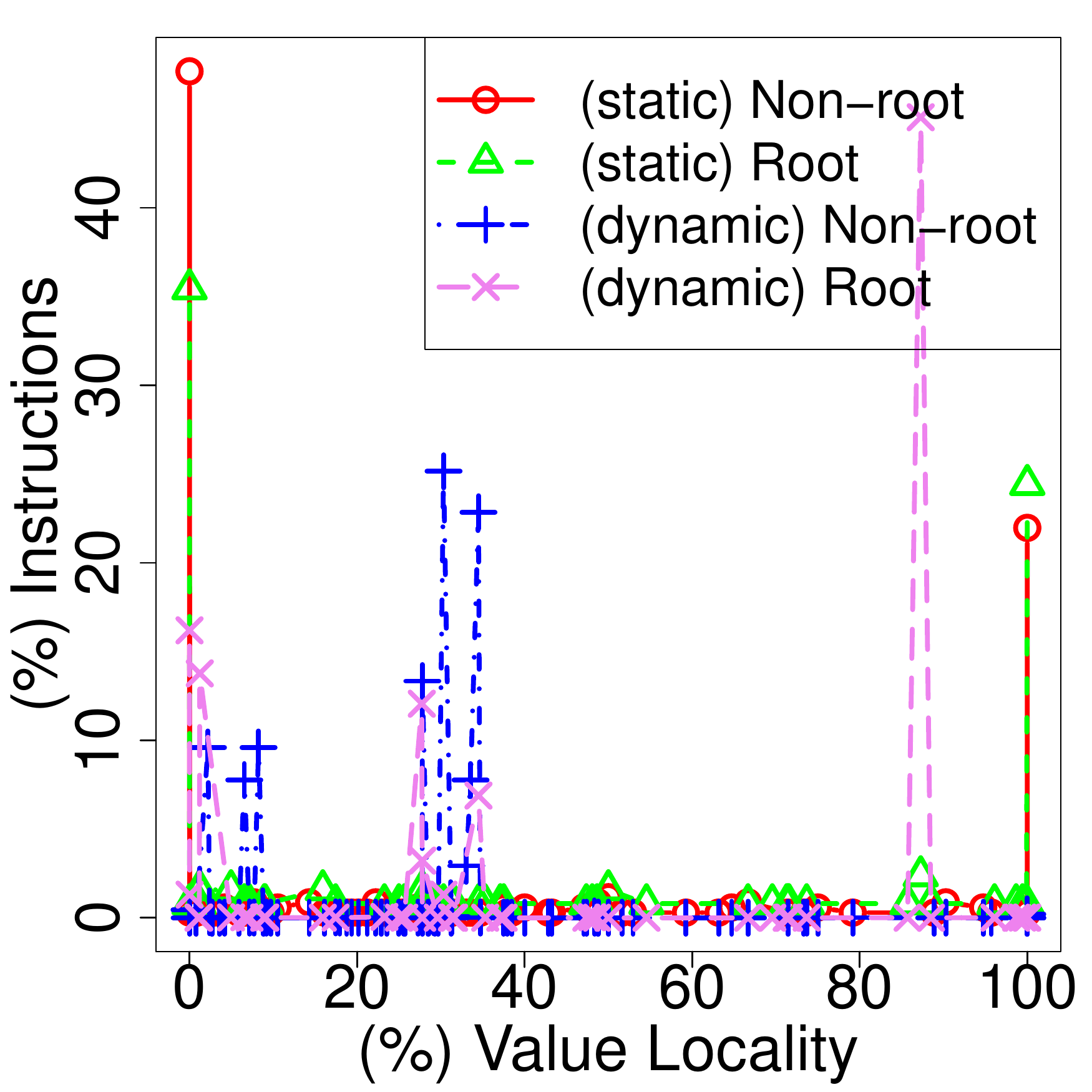}
        } \kern-0.8em
        \subfloat[ferret]{
                \includegraphics[width=0.25\textwidth]{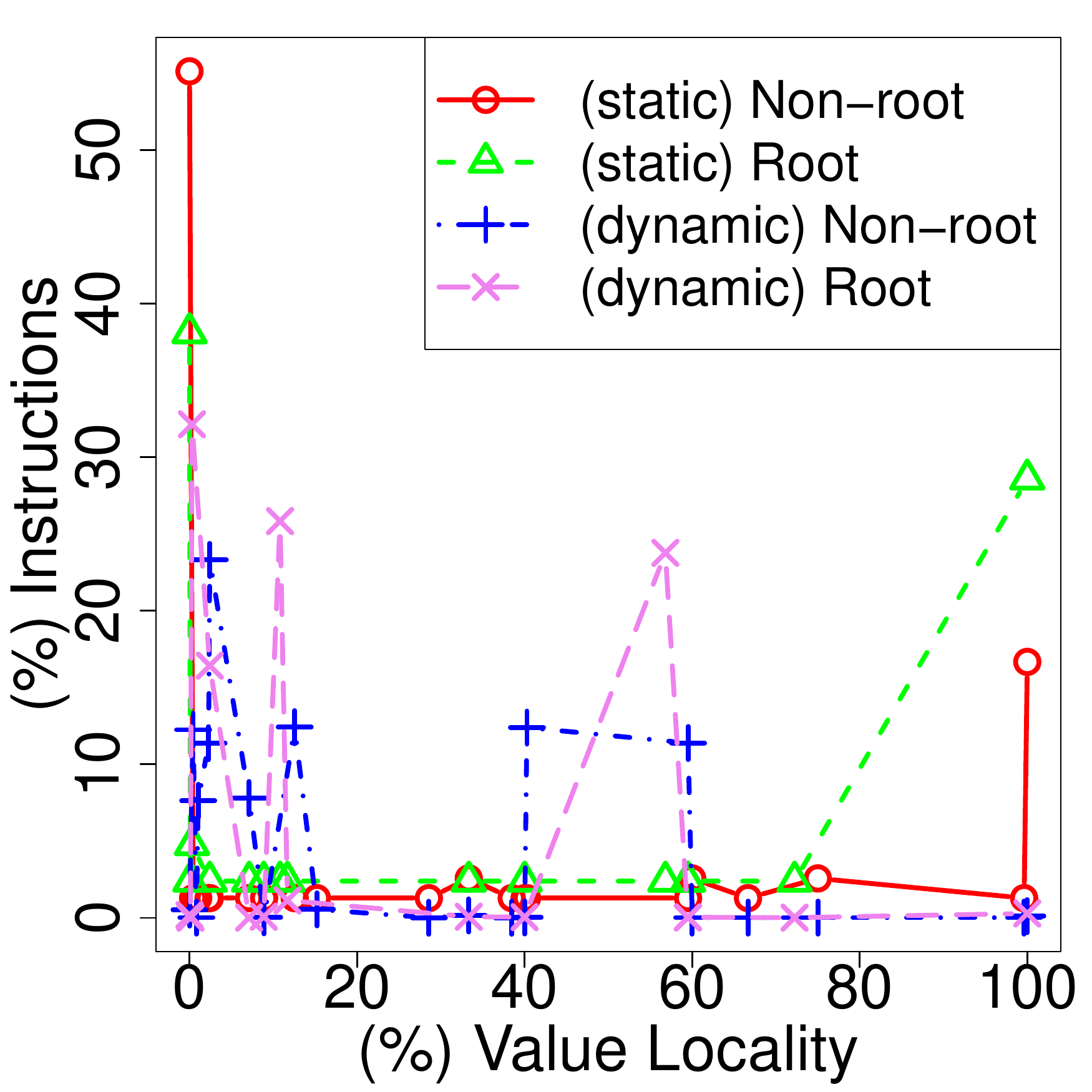}
        } \kern-0.8em
        \subfloat[raytrace]{
                \includegraphics[width=0.25\textwidth]{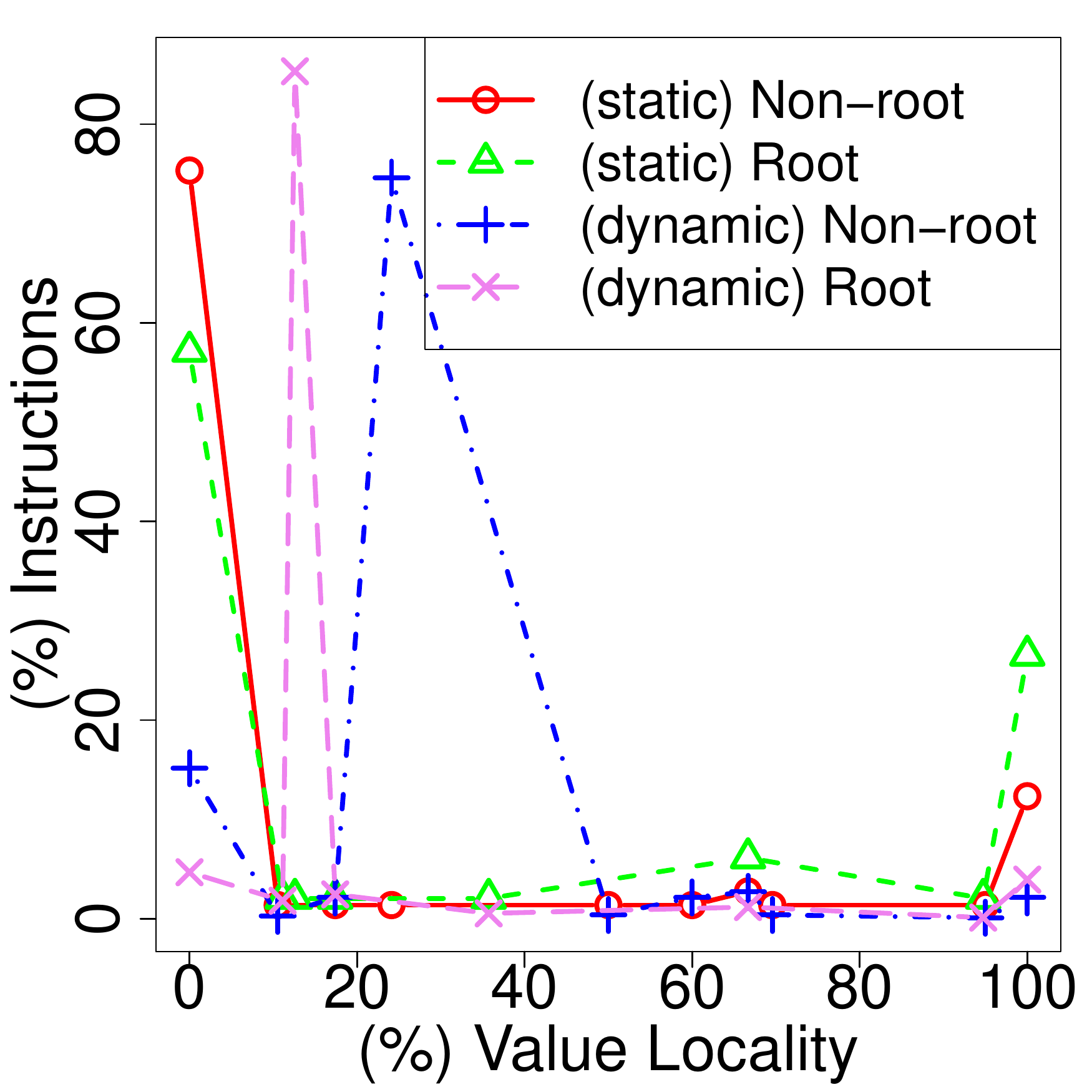}
        } \kern-0.8em
        \subfloat[backpropagation]{
                \includegraphics[width=0.25\textwidth]{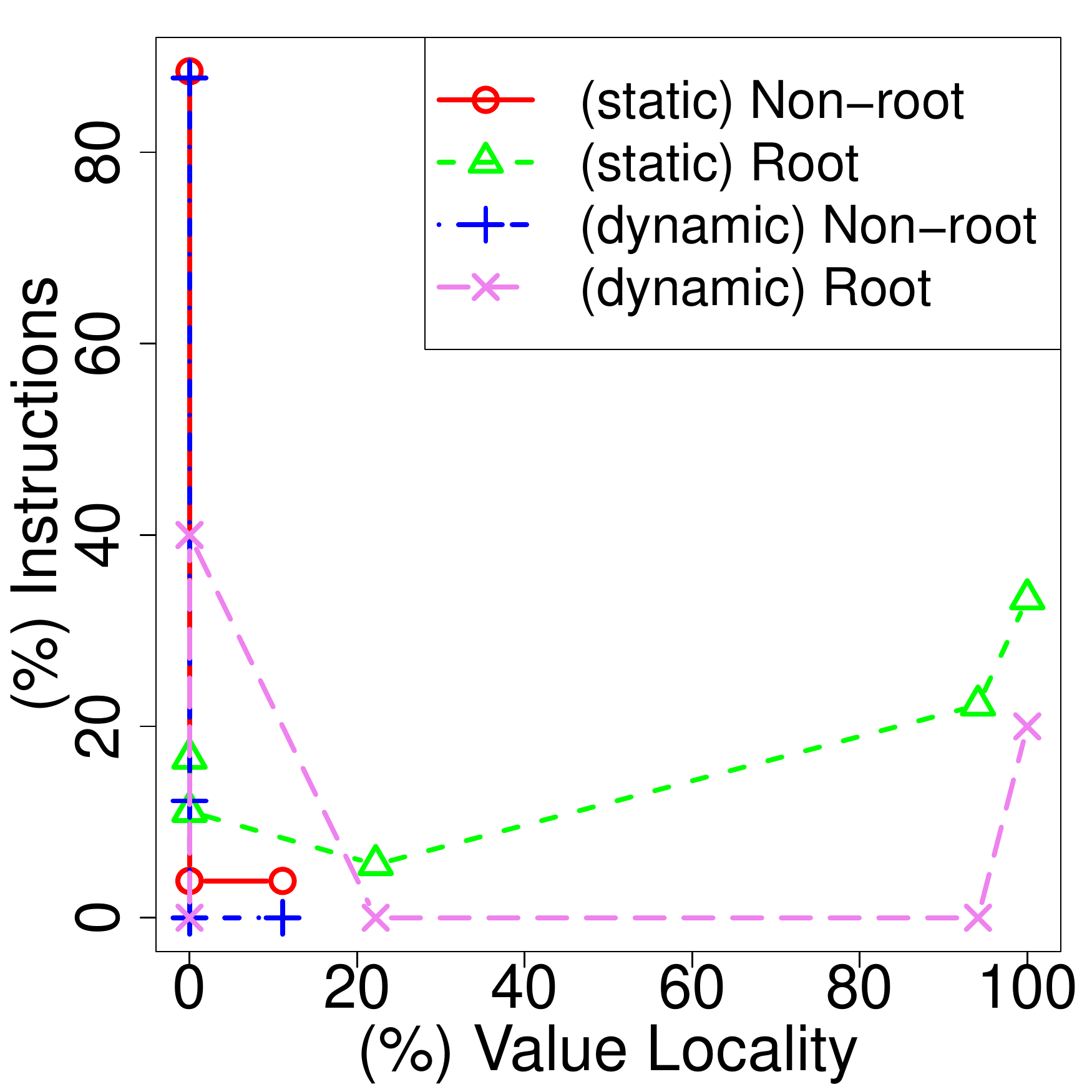}
                \label{fig:bp_locality}
        } \\ 
        \subfloat[bfs]{
                \includegraphics[width=0.25\textwidth]{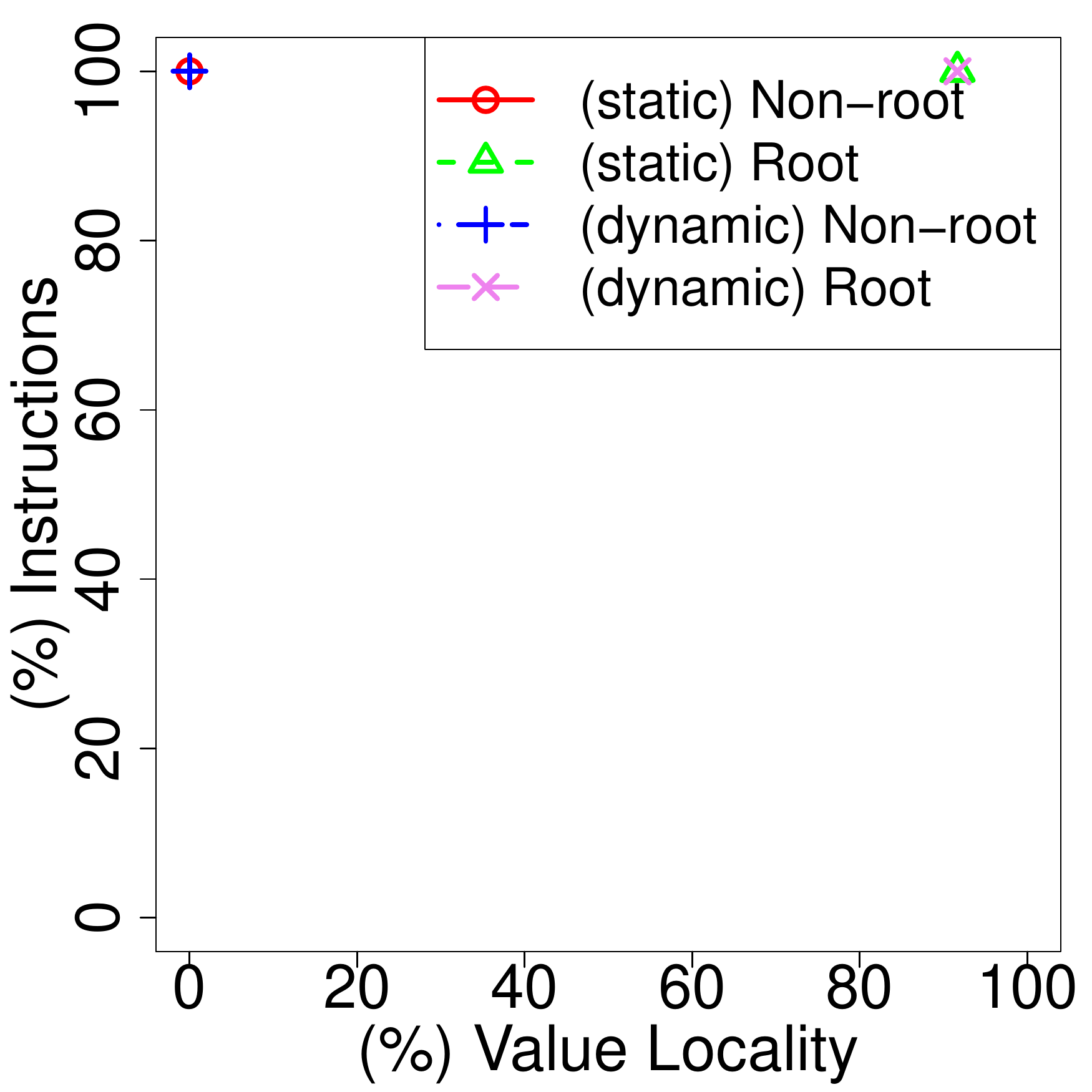}
        } \kern-0.8em      
        \subfloat[srad]{
                \includegraphics[width=0.25\textwidth]{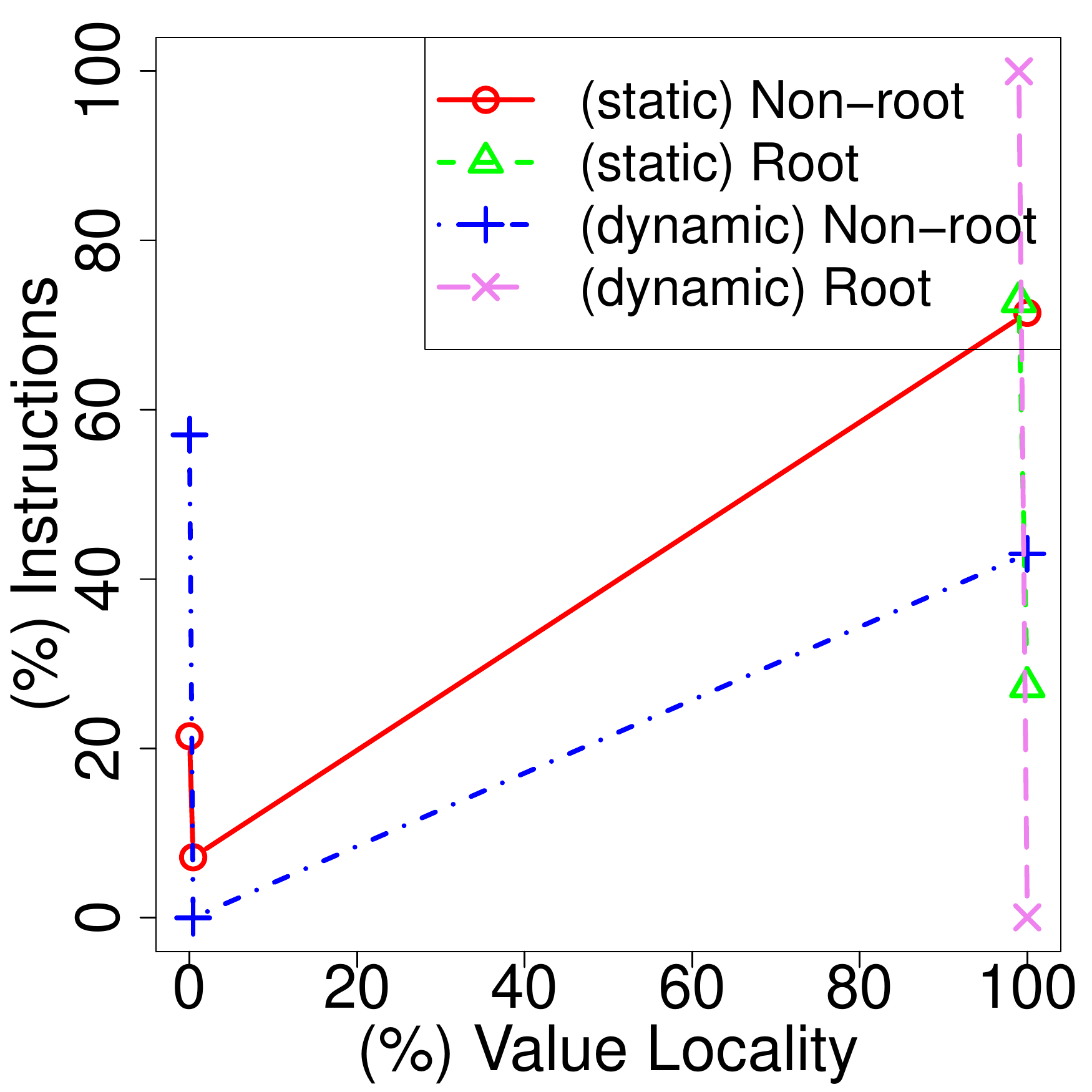}
		\label{fig:sr_locality}
        }
        \caption{Value locality of RSlice instructions.
  \label{fig:bb_locality}}
  \end{center}
\end{figure*}


\subsection{Impact on Execution Semantics}
\label{sec:eval2}
\begin{figure*}[h!]
  \begin{center}
	\kern-0.8em
        \subfloat[mcf]{
                \includegraphics[width=0.25\textwidth]{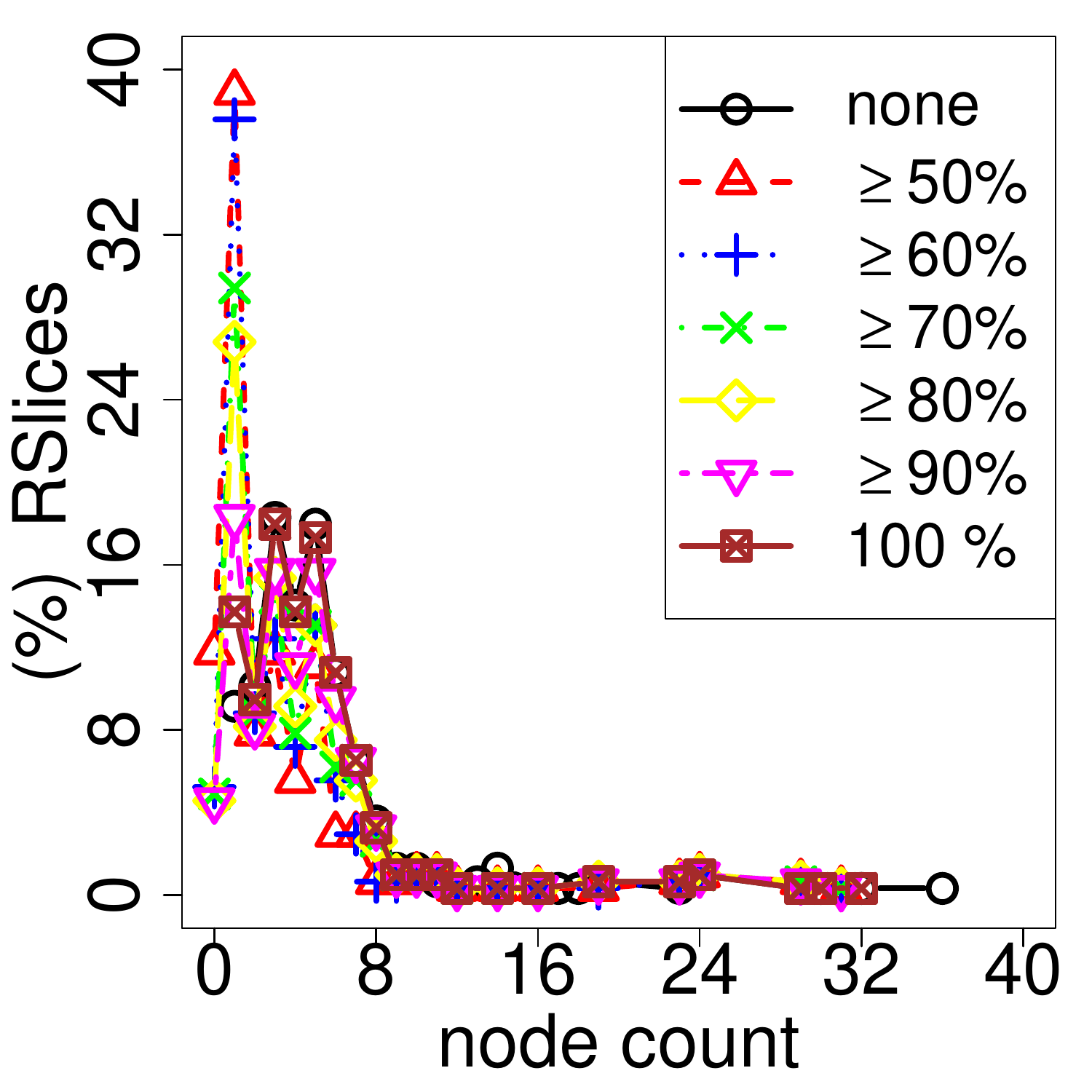}
        } \kern-0.8em
        \subfloat[sphinx3]{
                \includegraphics[width=0.25\textwidth]{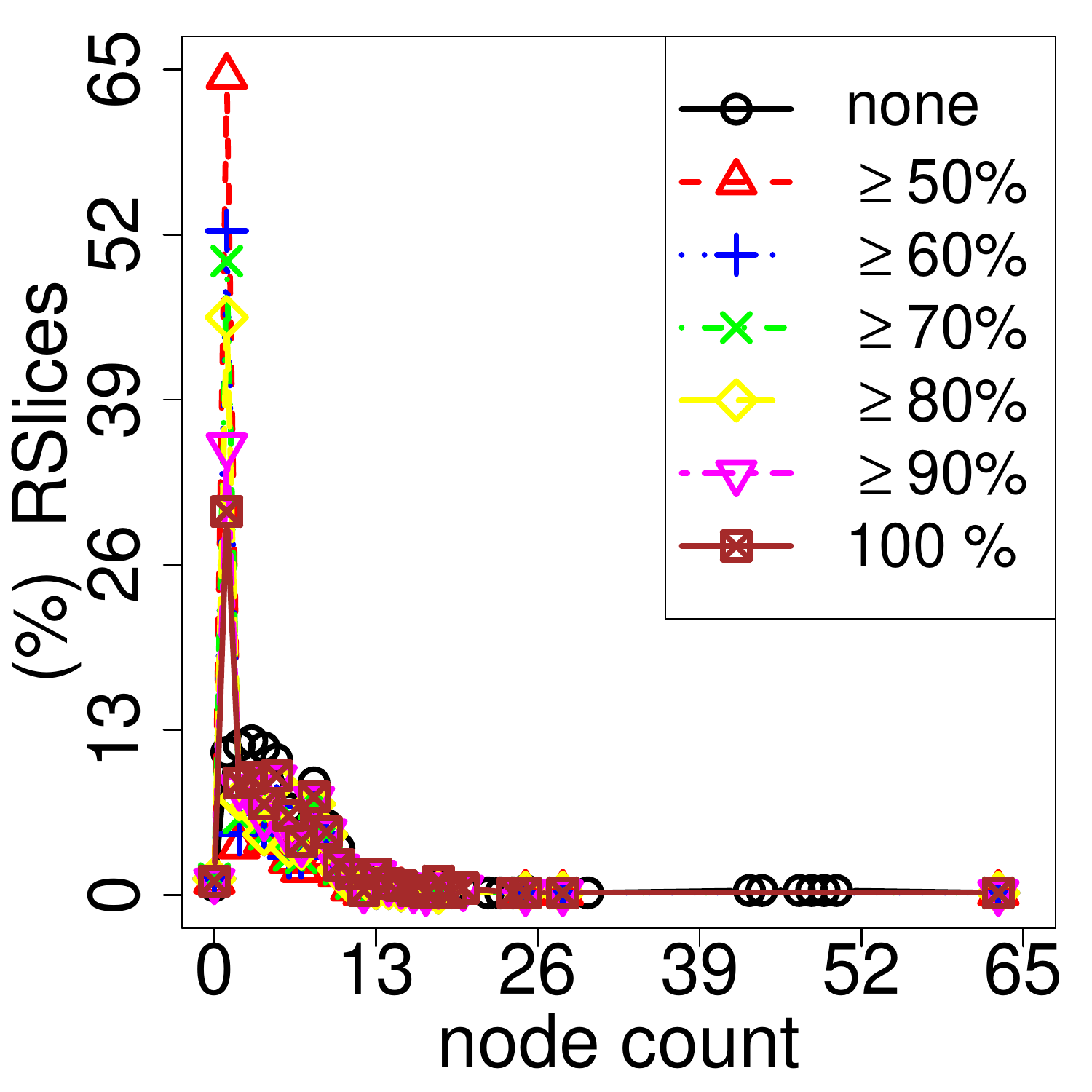}
        } \kern-0.8em
        \subfloat[is]{
                \includegraphics[width=0.25\textwidth]{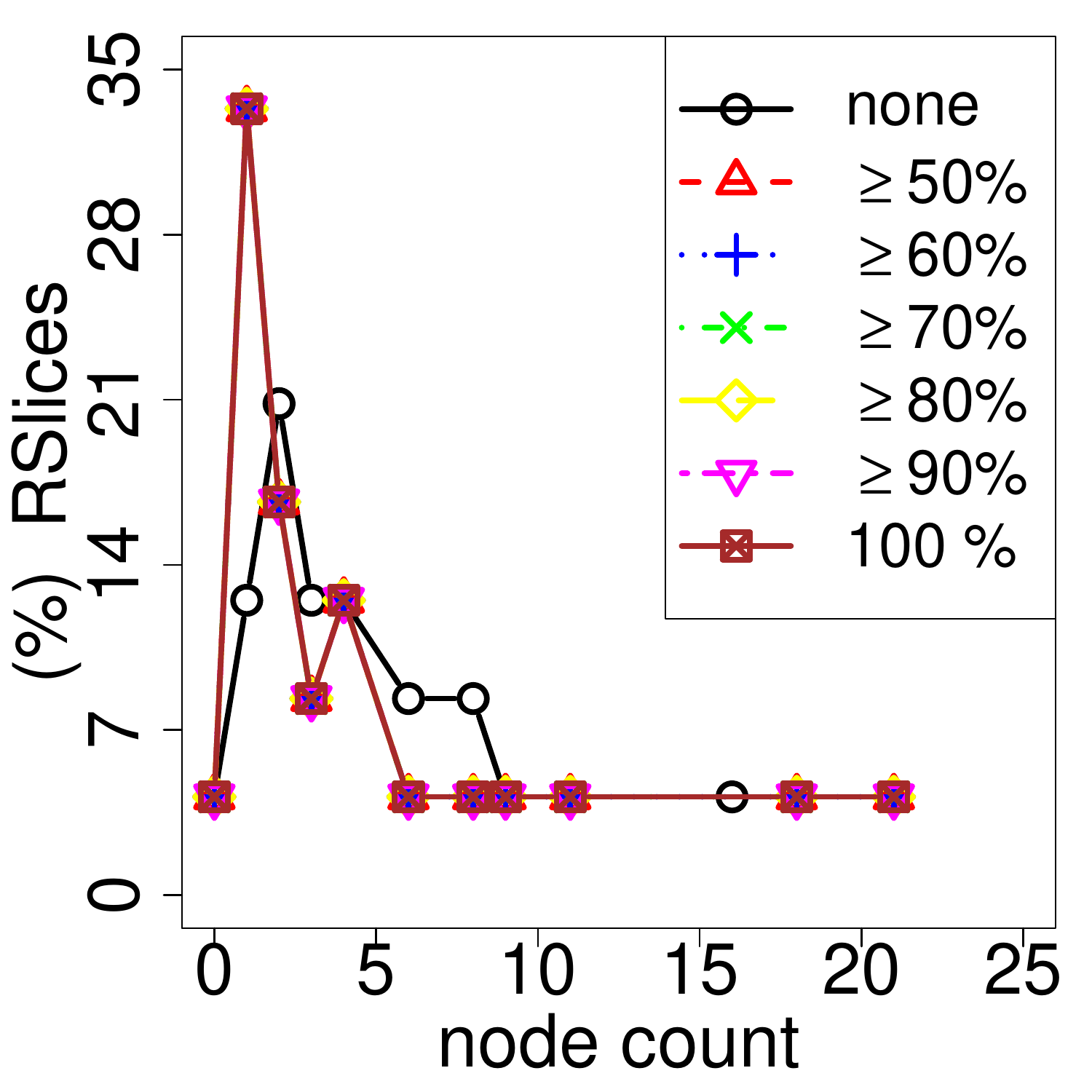}
        } \kern-0.8em
        \subfloat[canneal]{
                \includegraphics[width=0.25\textwidth]{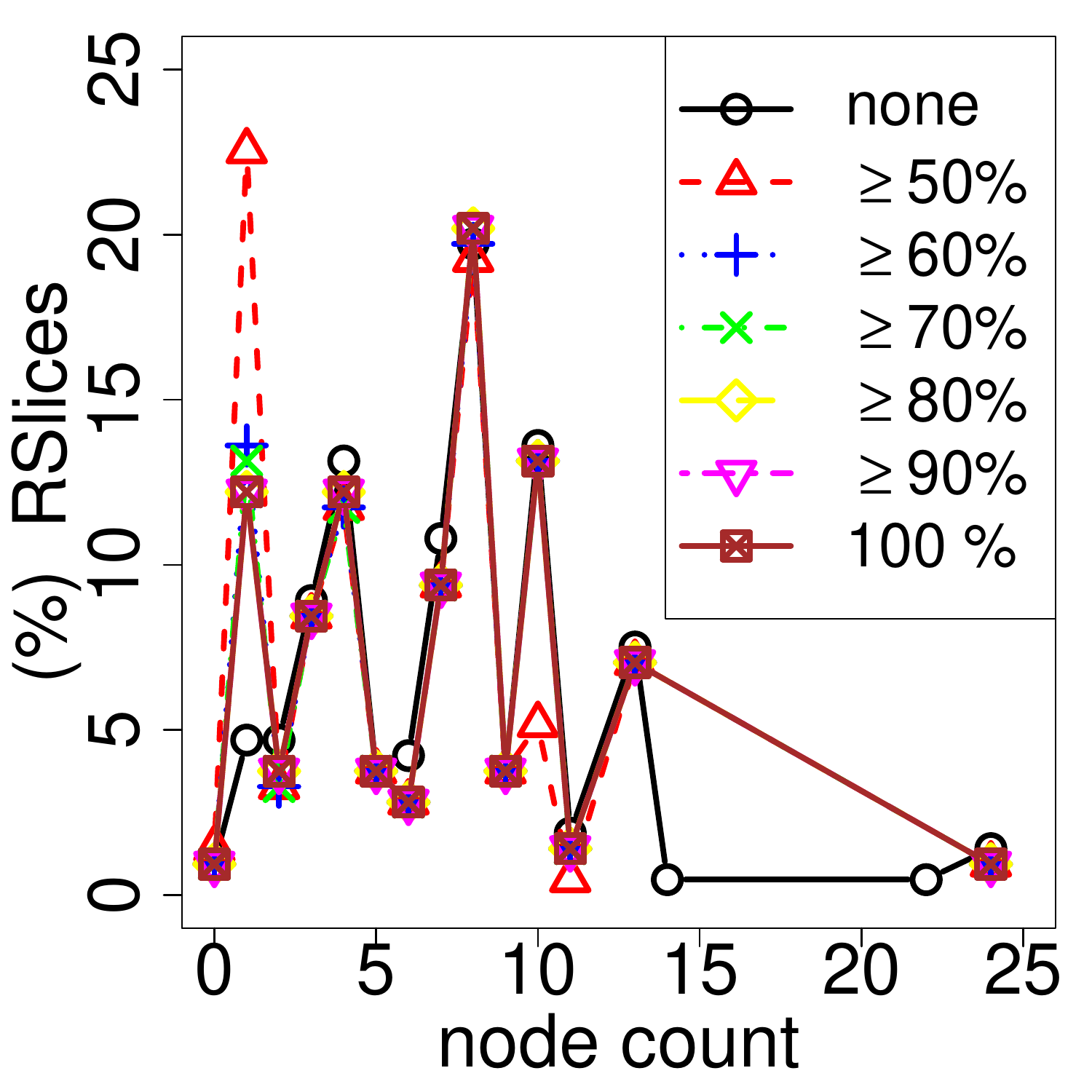}
        } \\ \kern-0.8em
        \subfloat[facesim]{
                \includegraphics[width=0.25\textwidth]{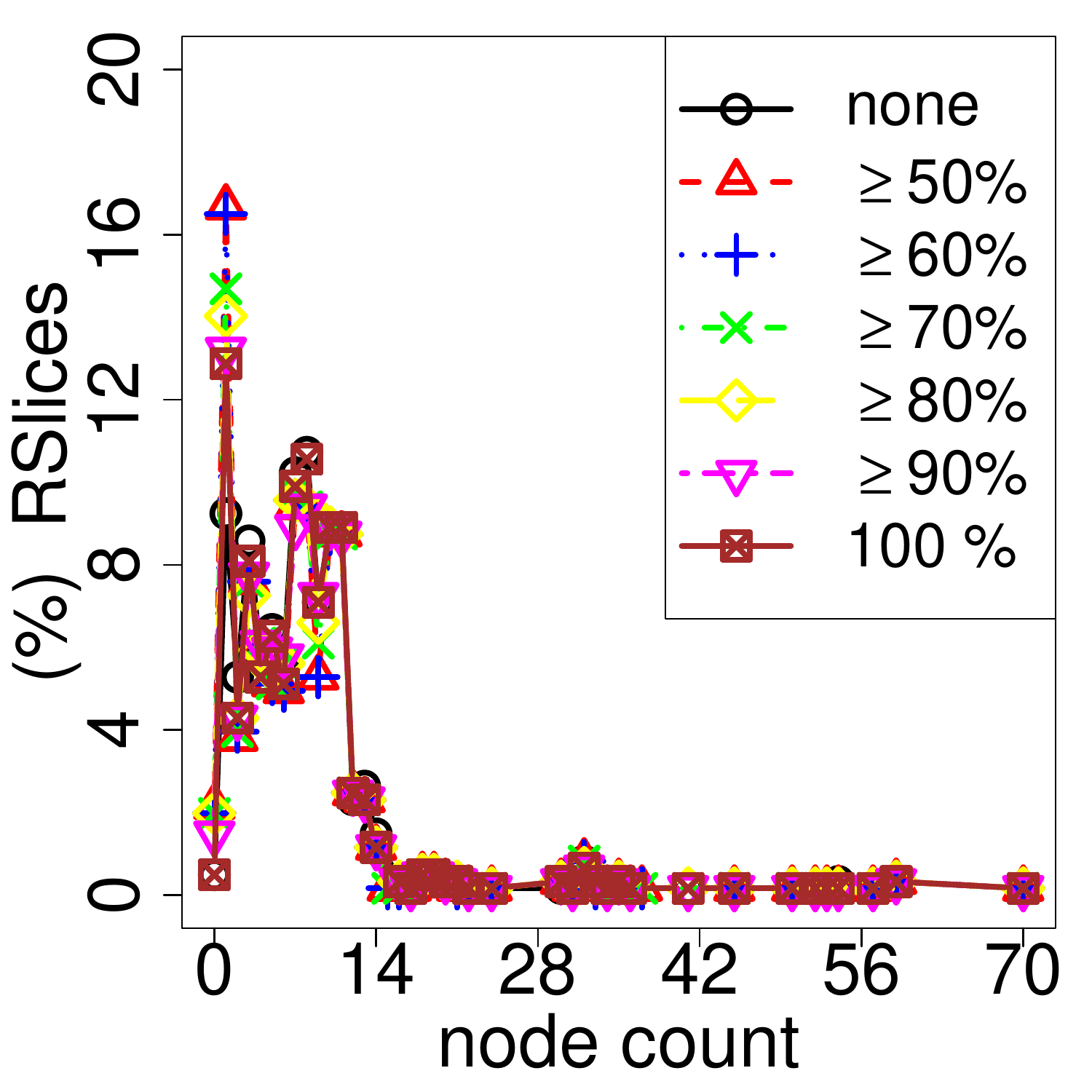}
        } \kern-0.8em
        \subfloat[ferret]{
                \includegraphics[width=0.25\textwidth]{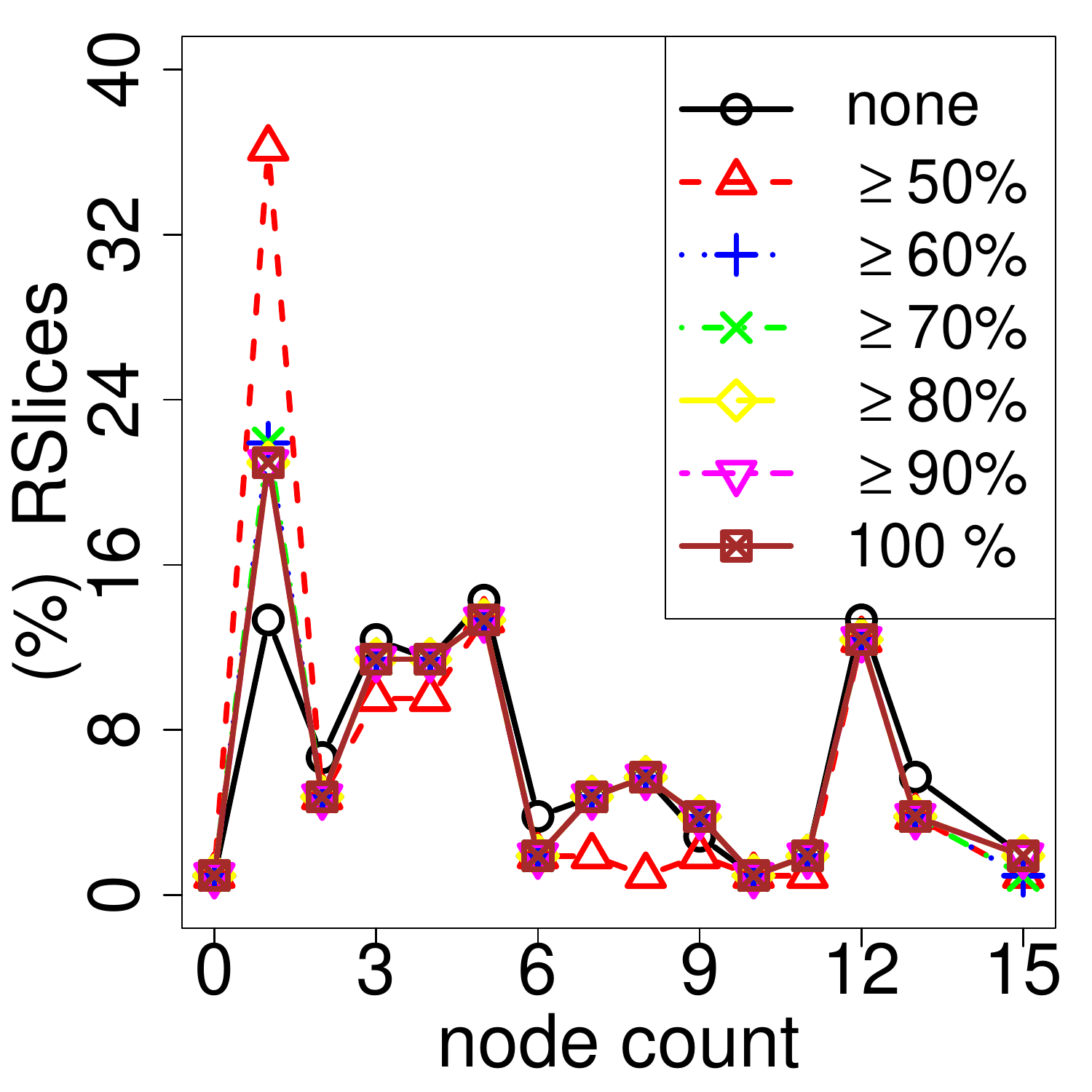}
        } \kern-0.8em
        \subfloat[raytrace]{
                \includegraphics[width=0.25\textwidth]{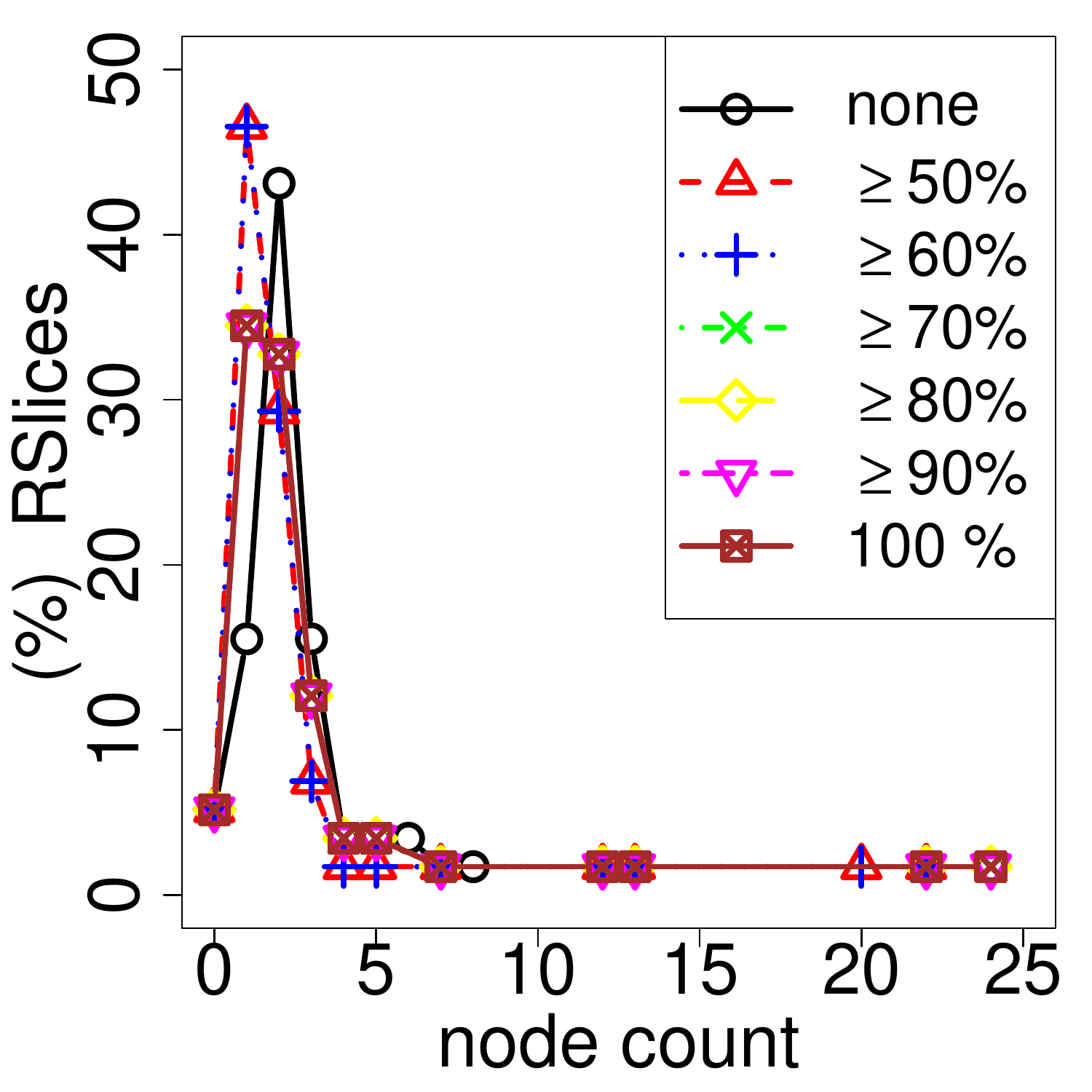}
        } \kern-0.8em
        \subfloat[backpropagation]{
                \includegraphics[width=0.25\textwidth]{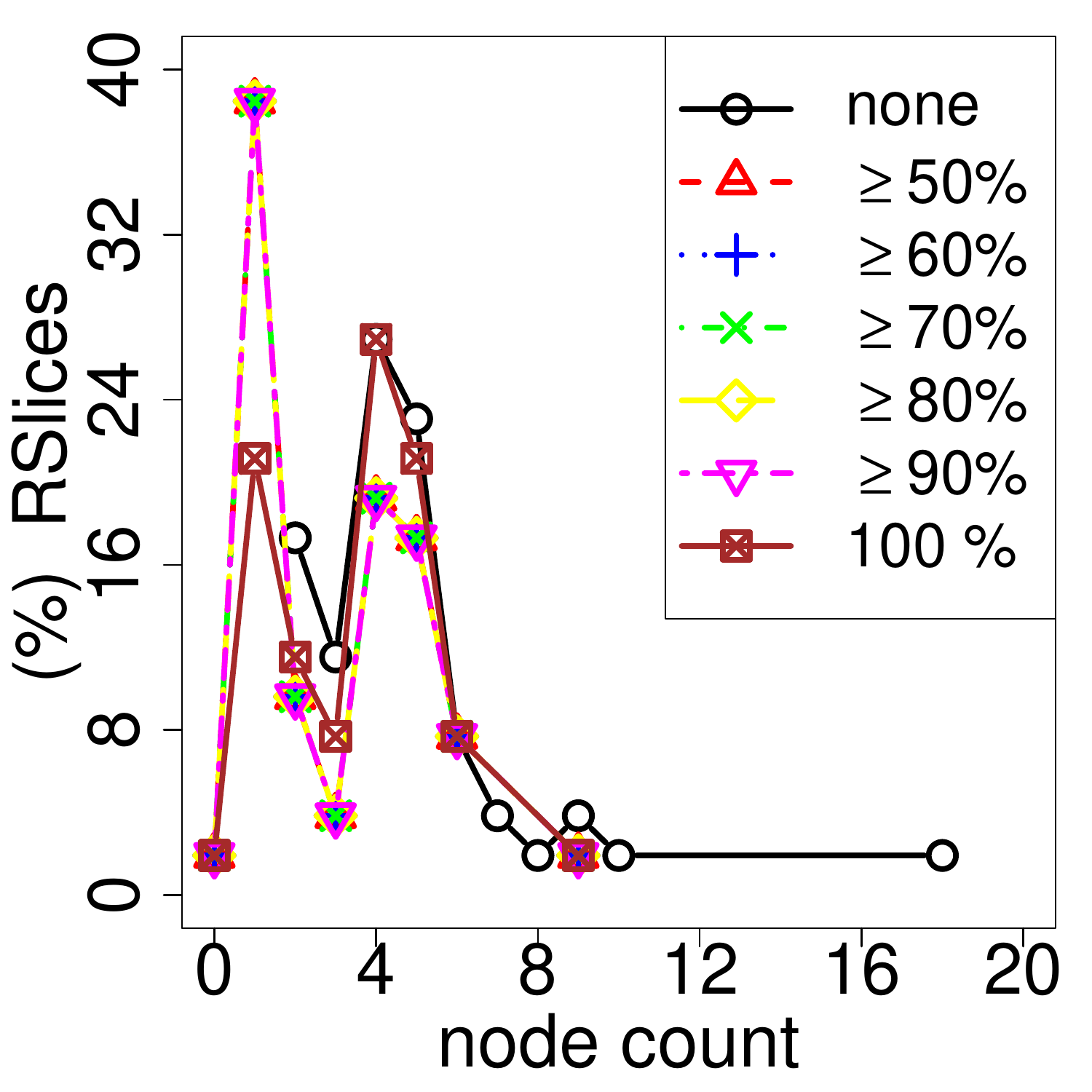}
        } \\
        \subfloat[bfs]{
                \includegraphics[width=0.25\textwidth]{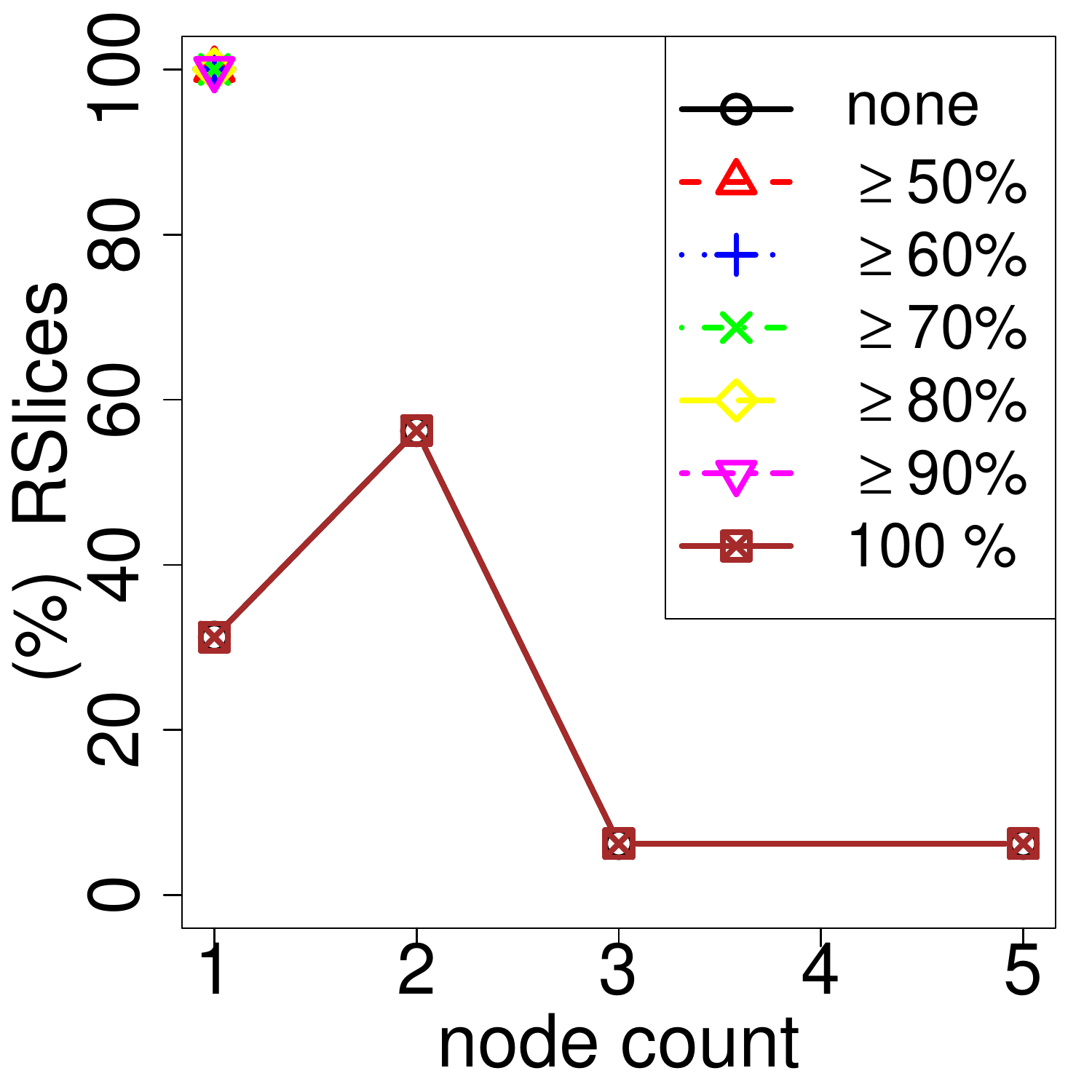}
        } \kern-0.8em
        \subfloat[srad]{
                \includegraphics[width=0.25\textwidth]{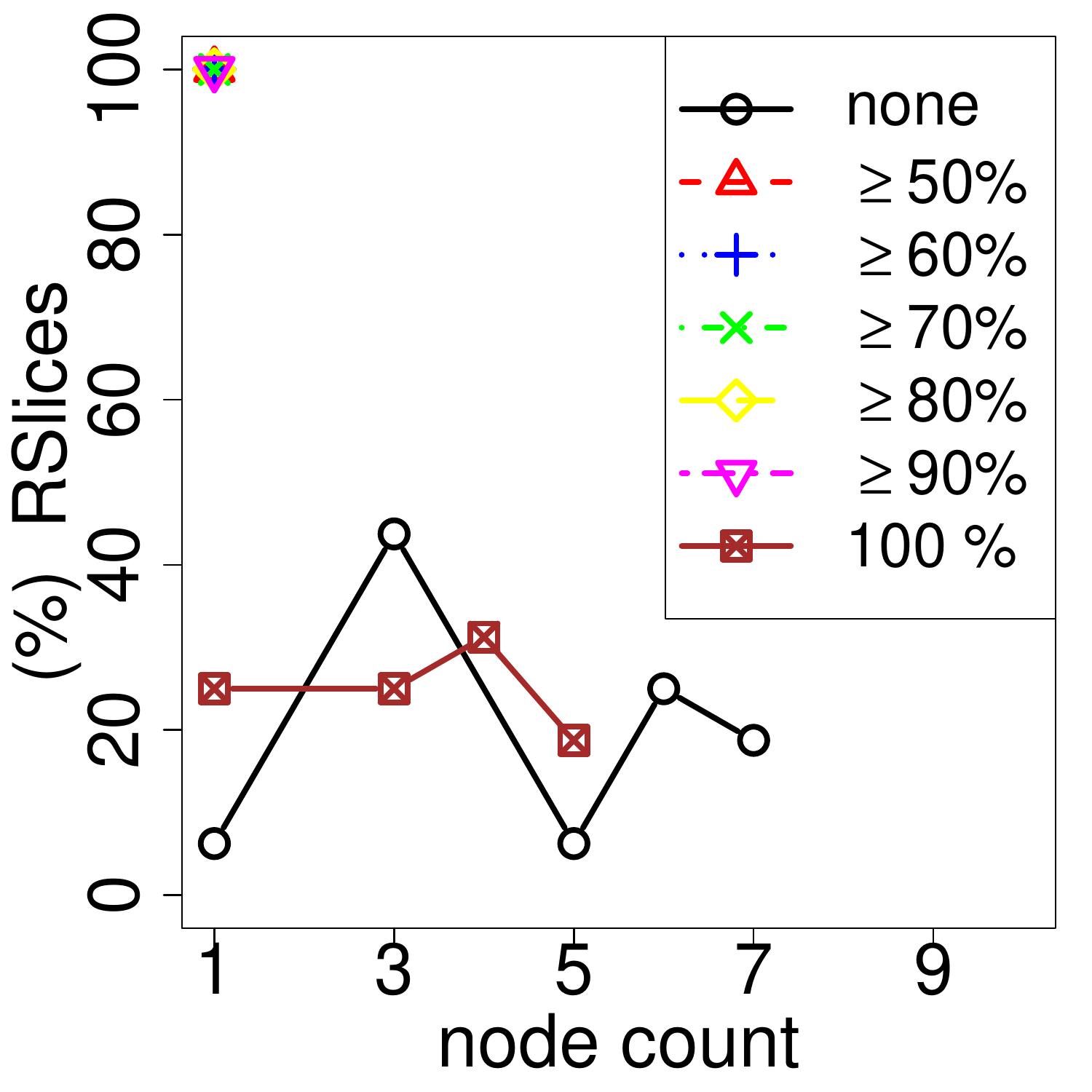}
        }
	\caption{Node count of RSlices before ({\bf recalculation}) and after pruning
		  ({\bf recalculation+prediction}). 
  \label{fig:rtree_height}}
  \end{center}
\end{figure*}

\noindent As explained in
Sections~\ref{sec:pred} and~\ref{sec:rPred}, in the context of recomputation, prediction serves two purposes:
\newline \noindent (i) to predict the values which would otherwise be loaded
from memory (and which
correspond to the values to be re-produced by RSlice roots
under pure {\bf recalculation}) under {\bf
prediction};
\newline \noindent (ii) to predict the input
values of intermediate (non-root) RSlice nodes under {\bf
recalculation+prediction}. 

{\bf Prediction} can eliminate re-execution along an entire RSlice if the values
to be re-produced by the RSlice root (i.e., the values which would otherwise be
loaded from memory) exhibit sufficiently high locality.
{\bf recalculation+prediction}, on the other hand, can prune any intermediate
RSlice node (except the root) exhibiting sufficient
(input) value locality to render a 
smaller RSlice, which in turn becomes less energy costly to execute. 

For prediction based recomputation to
work, the respective instructions should exhibit sufficiently high value locality.
Figure~\ref{fig:bb_locality} reports a histogram of \% value locality
(x-axis) for all instructions residing in RSlices. The y-axis reports the \% share of
instructions exhibiting a given value of locality on the x-axis.  {\em Root}
captures the output value locality of RSlice roots; {\em Non-root}, the
input value locality of intermediate (non-root) RSlice nodes. Recall that the
output value locality of RSlice roots corresponds to the value locality of
the respective load instructions which are replaced by RSlices.

%
Notice the distinction between static and dynamic instructions (for both root and
non-root, i.e., intermediate instructions).  Static instructions are the ones that are embedded in
the binary by the compiler. Dynamic instructions are the ones that are actually
executed at runtime. A static instruction may have multiple dynamic instances
executed at runtime, or may not be executed at all. This distinction helps us to
explain why, for instance, we do not obtain much benefit from {\bf prediction}
although a great fraction of static instructions have high value locality
for {\it is} (Figure~\ref{fig:is_locality}): 
38.46\% of (static) root instructions of {\it is} have 100\% value locality, but
{\it is} does not benefit much from {\bf prediction} (Figure~\ref{fig:pred_edp}).
This is because, at runtime, the root instructions having 100\% value locality are
not executed as many times as other root instructions that have lower value locality.  In
fact, less than 1\% of dynamic root instructions executed have 100\% value locality for
{\it is}, as shown in Figure~\ref{fig:is_locality}.  
The previous section revealed that bp benefits from {\bf prediction}
the most (Figure~\ref{fig:pred_edp}).  Therefore, we expect a larger fraction of
roots to have very high value locality for this benchmark.
Figure~\ref{fig:bp_locality} reveals that 20\% of dynamic root instructions
of bp have 100\% value locality indeed.  
A similar trend holds for
non-root instructions under {\bf recalculation+prediction}. For {\bf
recalculation+prediction}, prediction of non-root instructions can provide sizable gains
only if the dynamic share of non-root instructions exhibiting high value locality is
large.

Figure~\ref{fig:rtree_height} shows how the node count of RSlices change as the
locality threshold to enable prediction increases from 50\% to 100\% under {\bf
recalculation+prediction} -- {\em none} reflects no prediction, i.e., pure {\bf
recalculation}.
The figure reports a histogram of node count of RSlice (x-axis).
The y-axis reports the \% share of
RSlices having a given node count on the x-axis. 
A lower threshold enables more predictions, hence
more producer instructions can get
pruned, and the node count shrinks more. 
We observe that prediction at a value locality threshold of 50\% can reduce the node count of
RSlices up to 56\%.  

\section{Related Work}
\label{sec:rel}
\noindent 
Amnesiac~\cite{amnesiac17} trades off {computation for
communication} by replacing energy hungary loads with a set of low-energy arithmetic/logic instructions that are responsible for generating data 
to be loaded. This reduces the amount of energy consumed on data communication. We use similar compiler-based proof-of-concept recalculation implementation.
Kandemir et al.
proposed recomputation to reduce off-chip memory area in embedded
processors~\cite{Kandemir:2005gw}. Koc et al. investigated how recomputation of
data residing in memory banks in low-power states can reduce the energy
consumption by preventing frequent switching of the corresponding banks to
high-power states for data retrieval~\cite{Koc:2006ce}.  Koc et al. further
devised recomputation-aware compiler optimizations for scratchpad
memories~\cite{Koc:2007bl}. 
The compiler strategies from~\cite{Koc:2006ce} and~\cite{Koc:2007bl} are
confined to array variables.
In our paper, recomputation is not limited to embedded
processors or specific data
structures. 
DataScalar~\cite{datascalar} trades off {computation for
communication} by replicating data in each processor's local memory in a
distributed environment.
Accordingly, Datascalar divides the program address space between replicated and
communicated pages. Our framework {\em trades off computation for storage},
hence  minimizes communication rather as a side effect. As opposed to
DataScalar, our framework can reduce the program memory footprint.
Our study analyzes recomputation at a finer granularity.
Processing in/near memory~\cite{lim,pim,iram,flexram,petaflop} 
can bridge the gap between logic and memory speeds by embedding
compute capability in/near memory.
Processing in memory can minimize energy-hungry 
data transfers, as well, and is orthogonal to recomputation. 
Memoization~\cite{Sodani:1997hn,resistiveComp} -- the dual of recomputation --
replaces (mainly frequent and expensive) computation with table look-ups for
pre-computed data. Similar to processing in memory and recomputation,
memoization can mitigate the communication overhead (as long as table look-ups
remain cheaper than long-distance data retrieval). Memoization and recomputation
can complement each other in boosting energy efficiency.

\section{Conclusion}
\label{sec:conc}
\noindent Recomputation can minimize, if not eliminate, the prevalent power and
performance (hence, energy) overhead incurred by data storage, retrieval, and
communication, thus, render more energy efficient execution.  
This paper provided a quantitative proof-of-concept analysis for the computation
vs. communication trade-off, along with a taxonomy.  Recomputation replaces data
load(s) from memory with the reproduction of the respective data.  Unless the
energy cost of reproducing data remains less than the energy cost of retrieving
it from memory, recomputation cannot improve energy efficiency.  

In this study,
we explored (interactions between) two broad classes of recomputation techniques: brute-force
recalculation and prediction based recomputation. Under recalculation, the recomputation effort goes
to the derivation of the data values (which would otherwise be loaded from
memory), by re-executing the producer instruction(s) of the eliminated load(s).
Under prediction, the recomputation effort goes to the estimation of the data
values by exploiting value locality -- the likelihood of the recurrence of
values (which would otherwise be loaded from memory) within the course of
execution.  We find that recalculation has wider coverage for recomputation than
prediction, as prediction cannot be effective under limited value locality.

\bibliographystyle{plain}
\bibliography{referencesShortLabel,amnesiac}

\end{document}